\documentclass[journal=jctcce,manuscript=article]{achemso}

\usepackage{hyperref}

\usepackage{multirow}
\usepackage{graphicx}

\usepackage{adjustbox}

\usepackage[version=4,arrows=pgf-filled,
]{mhchem}
\usepackage{caption}
\usepackage[labelformat=parens, font=sf]{subcaption}
\usepackage{comment}
\usepackage{xparse}
\usepackage{multirow, array}
\usepackage{tabularx,ragged2e,booktabs, array}
\usepackage{siunitx}
\DeclareSIUnit\uc{\protect \text UC}
\DeclareSIUnit\angstrom{\protect \text {Å}}
\usepackage[nolist, nohyperlinks]{acronym}
\usepackage{xcolor}
\usepackage{enumitem}
\usepackage{svg}
\usepackage{amsmath}

\usepackage{listings}
\selectcolormodel{RGB}

    \definecolor{bluebright}{RGB}{0, 196, 223}
    \definecolor{bluemuted}{RGB}{0, 79, 113}
    \definecolor{blueuniversity}{RGB}{4, 30, 66}
    \definecolor{reduniversity}{RGB}{193, 0, 67}
    \definecolor{brown}{RGB}{106, 51, 40}
    \definecolor{brownmuted}{RGB}{109, 79, 71}
    \definecolor{pinkbright}{RGB}{208, 0, 111}
    \definecolor{pinkmuted}{RGB}{184, 133, 141}
    \definecolor{orangebright}{RGB}{205, 90, 19}
    \definecolor{yellowbright}{RGB}{244, 170, 0}
    \definecolor{yellowmuted}{RGB}{156, 154, 0}
    \definecolor{greenbright}{RGB}{41, 188, 41}
    \definecolor{greendark}{RGB}{21, 71, 52}
    \definecolor{redbright}{RGB}{172, 0, 64}
    \definecolor{blue}{RGB}{0, 114, 136}
    \definecolor{bluelight}{RGB}{194, 211, 223}
    \definecolor{purple}{RGB}{131, 0, 101}
    \definecolor{turquoisemuted}{RGB}{69, 126, 129}
    
    \definecolor{URsemi}{gray}{0.55}
    \definecolor{URcitation}{named}{greendark}
    \definecolor{URurl}{named}{blueuniversity} 
    \definecolor{URtitle}{named}{blueuniversity}
    \definecolor{URlink}{named}{blueuniversity} 
    \definecolor{halfgray}{gray}{0.55}
    \definecolor{webgreen}{named}{greendark}
    \definecolor{webbrown}{named}{brown}
    \colorlet{clay}{orangebright!50!yellowbright}
    \colorlet{il}{bluelight!30!turquoisemuted!60!greenbright!70!white}
    \colorlet{bulk}{bluelight!90!blueuniversity!60!bluebright!40!}

\usepackage{multicol}

\usepackage{tikz}

\usetikzlibrary{calc, er, snakes, trees, backgrounds, arrows, shapes.geometric, positioning, patterns, matrix, decorations.pathreplacing, calligraphy, fit, 3d}
\usepackage{pgf}
\usepackage{pgfplots}
\usepackage{pgfbaselayers}
\usepackage{pgfmath}
\pgfdeclarelayer{background}
\pgfdeclarelayer{foreground}
\pgfsetlayers{background, foreground}
\pgfplotsset{compat=1.8}
\tikzstyle{arrow} = [thick,->,>=stealth, shorten <= 3 mm, shorten >= 3 mm]
\tikzstyle{line} = [thick]
\tikzstyle{head} = [thick,->,>=stealth]

                \tikzstyle{rect} = [rectangle,
                                    rounded corners,
                                    thick,
                                    color=black,
                                    inner ysep=0.02\textwidth,
                                    ]
                                    
                \tikzstyle{stepnum} = [thick,
                                        color=black, 
                                        inner ysep=0.02\textwidth,
                                        outer ysep=0.03\textwidth,
                                        draw=none,
                                        ]
                \tikzstyle{desc} = [rounded corners,
                                    draw=black,
                                    outer ysep=0.03\textwidth,
                                    inner sep=0.02\textwidth,
                                    minimum height=1, 
                                    thick,
                                ]                
              \tikzstyle{sep} = [draw=black,
                                dashed,
                                thick,
                                shorten <= -0.25\textwidth
                                ]
            \tikzstyle{flowarrow} = [-{triangle 60}, thick]

\NewDocumentCommand{\createAxisfig}{s m t' O{x=0, y=0} r{i}{,} d{+}{,} D{!}{,}{south west} O{x=1, y=1} }{
\resizebox{\textwidth}{!}{
    \begin{tikzpicture}
            \IfBooleanTF{#3}{\def\stransp{0}}{\def\stransp{1}}
            \begin{scope}[scale=1, x=\textwidth, y=\textwidth, z=-0.385\textwidth, every path/.style={-{Triangle[angle=45:1pt 3, scale=2]}, thick, cap=round, line join=round, miter limit=25, line width=1, cap=round, draw=black, opacity=\stransp,}, shape rectangle/.style={draw=none}, every node/.style={circle, draw=none, font=\sffamily\small, outer sep=1pt, minimum width=1em, minimum height=1ex, inner sep=0}]
            \pgfkeys{
            /names/.cd,
            a/.initial = 1,
            b/.initial = 2,
            c/.initial = 3,
            }
            \pgfkeys{/names/.cd, #2}
            \pgfkeys{
            /shifts/.cd, 
            x/.initial = 0,
            y/.initial = 0,
            }
            \pgfkeys{/shifts/.cd, #4}
            \pgfkeys{
            /flip/.cd, 
            x/.initial = 1,
            y/.initial = 1,
            }
            \pgfkeys{/flip/.cd, #8}
                \begin{pgfonlayer}{background}
                    \pgfmathsetmacro{\xflip}{\pgfkeysvalueof{/flip/x}}
                    \pgfmathsetmacro{\yflip}{\pgfkeysvalueof{/flip/y}}
                    \node (clay) [rectangle, anchor=#7, opacity=1] {\scalebox{\xflip}[\yflip]{\includegraphics[width=\textwidth]{#5}}};
                \end{pgfonlayer}
                \begin{pgfonlayer}{foreground}
                    \IfBooleanTF{#1}{\def\origin{north west}}{\def\origin{south west}}
                    \IfBooleanTF{#1}{\def\invert{-}}{\def\invert{}}
                    \IfBooleanTF{#1}{\def\banchor{south east}}{\def\banchor{north west}}
                    \IfBooleanTF{#1}{\def\canchor{south}}{\def\canchor{north}}
                    \pgfmathsetmacro{\yshift}{\pgfkeysvalueof{/shifts/x}}
                    \pgfmathsetmacro{\xshift}{\pgfkeysvalueof{/shifts/y}}
                    \IfNoValueTF{#6}{\def\linestyle{solid}\def\depth{}}{\def\linestyle{#6}\def\depth{-}\IfBooleanF{#1}{\def\banchor{south}}}
                    \begin{scope}[canvas is xy plane at z=0]
                    \draw [] ($ (clay.\origin) + (\yshift, \xshift) $ ) -- ++  (0, \invert0.65)  node [pos=0.5, anchor=east, align=center] {${\pgfkeysvalueof{/names/c}}$};
                    \draw [overlay] ( $ (clay.\origin) + (\yshift, \xshift) $) -- ++( 0.65, 0) node [pos=0.5, anchor=\canchor, align=center] {${\pgfkeysvalueof{/names/a}}$};
                    \end{scope}
                    \draw [overlay, -{Triangle[angle=45:1pt 3,slant=-0.38,scale=2]}, \linestyle] ( $ (clay.\origin) + (\yshift, \xshift, 0) $) -- ++(0, 0, \depth0.35) node [pos=0.4, anchor=\banchor, align=center] {${\pgfkeysvalueof{/names/b}}$};
                \end{pgfonlayer}
            \end{scope}
        \end{tikzpicture}
    }
}

\DeclareSIUnit{\elementarycharge}{\text{\ensuremath{e}}}
\DeclareSIUnit{\uc}{\text{UC}}

\author{Hannah Pollak}
\affiliation[Chem]
{School of Chemistry, University of Edinburgh, Joseph Black Building, David Brewster Road, Edinburgh, EH9 3FJ, United Kingdom}

\author{Matteo T. Degiacomi}
\affiliation[Durham University]
{Department of Physics, Durham University, South Road, Durham, DH1 3LE, United Kingdom}

\author{Valentina Erastova}
\email{valentina.erastova@ed.ac.uk}
\affiliation[Chem]
{School of Chemistry, University of Edinburgh, Joseph Black Building, David Brewster Road, Edinburgh, EH9 3FJ, United Kingdom}
\alsoaffiliation[Astrobio]
{UK Centre for Astrobiology, School of Physics and Astronomy, University of Edinburgh, James Clerk Maxwell Building, Peter Guthrie Tait Road, Edinburgh, EH9 3FD, United Kingdom}

\title[ Modelling clays with ClayCode ]
  {Modelling realistic clay systems with ClayCode}

\abbreviations{}


\begin{document}

\begin{acronym}
    \acro{UC}{unit cell}
    \acro{MD}{molecular dynamics}
    \acro{AMCSD}{American Mineralogist Crystal Structure Database}
    \acro{T}{tetrahedral}
    \acro{O}{octahedral}
    \acro{diO}{dioctahedral}
    \acro{triO}{trioctahedral}
    \acro{SWy-123}{Wyoming montmorillonite}
    \acro{IMt-1}{Montana illite}
    \acro{KGa-1}{Georgia kaolinite}
    \acro{LDH}{layered double hydroxide}
    \acro{IS}{inner-hydration sphere}
    \acro{OS}{outer-hydration sphere}
    \acro{IL}{interlayer}
    \acro{SPC}{simple point charge}
\end{acronym}

\begin{tocentry}




  \begin{figure}[H]
    \centering
    \includegraphics{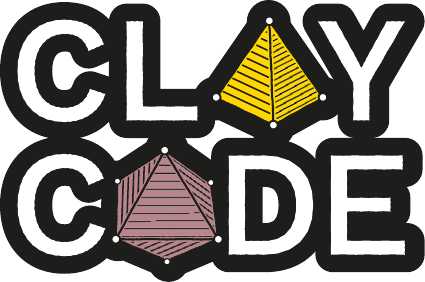}
    \label{fig:logo}
  \end{figure}
  
\end{tocentry}


\begin{abstract}
Clays are a broad class of ubiquitous layered materials. Their specific chemophysical properties are intimately connected to their molecular structure, featuring repeating patterns broken by substitutions. Molecular dynamics simulations can provide insight into the mechanisms leading to the emergent properties of these layered materials, however up to now idealised clay structures have been simulated to make the modelling process tractable. We present \textit{ClayCode}, software facilitating the modelling of clay systems closely resembling experimentally determined structures. By comparing a realistic model to a commonly used montmorillonite clay model, we demonstrate that idealised models feature noticeably different ionic adsorption patterns. We then present an application of \textit{ClayCode} to the study the competitive barium and sodium adsorption on Wyoming montmorillonite, Georgia kaolinite, and Montana illite, of interest in the context of nuclear waste disposal.
\end{abstract}

\section{Introduction}

Clay minerals are ubiquitous on Earth, and are also found on other rocky planetary bodies.
They are typically formed through prolonged chemical weathering of silicate-bearing rocks in the presence of water, or during a hydrothermal activity. Therefore, clays are a key component of soil and are crucial for various geological processes, including soil formation, weathering, and diagenesis. Furthermore, clay minerals play a central role in industrial and environmental applications, such as improved soil fertility, contaminant management, mining, and water purification. Clay mineral structure is affected by environmental conditions during their formation, which leads to a large variety of mineral structures with a broad range of properties. To predict the behaviour of clay minerals in changing environments or during an industrial application, we must understand the relationship between their structure and their properties. 

Molecular simulations have been developed for over the last half a century to offer atomistic-level insights into the structural and dynamic properties of molecular systems. While their application in the biomolecular area has been pervasive, their uptake into clay research has only started in the last 20 years with the development of specific sets of force field parameters suitable for clay simulations.\cite{Skipper1995a, Cygan2004, Heinz2013, Cygan2021}

\begin{figure}
    \centering
    \includegraphics[width=1.0\textwidth]{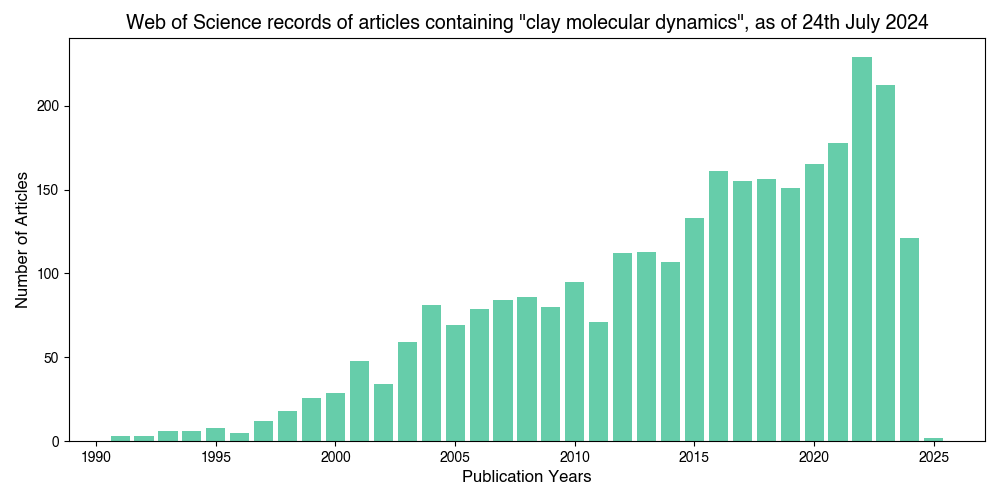}
    \caption{Number of research articles published per year mentioning "clay molecular dynamics", the record data is obtained from the Web of Science on the 24th July 2024.}
    \label{fig:publication_plot}
\end{figure}

While the number of publications utilising \ac{MD} simulations for clay studies is growing (see Figure \ref{fig:publication_plot}), the variety of clay models studied is still very limited and nohow representative of the broad array of naturally occurring clay minerals. This deficit can be only partially attributed to the narrow range of force field parameters currently available. Indeed, one of the key difficulties lies in the preparation of the clay model itself for the simulation, which is a time-consuming and error-prone procedure.

To facilitate the model setup and simulation of clay minerals representative of their natural counterparts, we have developed \emph{ClayCode} \href{https://github.com/Erastova-group/ClayCode}{github.com/Erastova-group/ClayCode}. This software enables the preparation of clay models closely matching their experimentally-determined counterparts, and the assignment of ClayFF force field \cite{Cygan2004, Cygan2021} parameters. All input files necessary for simulating the resulting parameterised model with GROMACS,\cite{Berendsen1995Gromacs} one of the fastest and most commonly used engines, are finally produced.

In this paper, after an overview of clay structures, we review the current state-of-the-art in the molecular modelling of clays, the tools available to assist users in the preparation of simulations, and their limitations. We then detail the \emph{ClayCode} workflow, highlighting how its flexibility and modularity enable its applicability to a broad range of molecular systems. To demonstrate the importance of modelling realistic clay systems, we compare adsorption of metal ions on two simulated models of Wyoming montmorillonite --- a common simple model previously used by us\cite{Underwood2015, Underwood2016} and other researchers\cite{Zhang2014, Zhang2016surface} ---
and a new model produced by \emph{ClayCode}, truthful to the experimentally characterised structure of the mineral.
Our results demonstrate that the subtle structural differences observed in natural clay minerals are key determinants of the mineral's physical properties.
Finally, we demonstrate the usage of \emph{ClayCode} in a real-case-scenario by comparing the adsorption of two metal ions, \ce{Na+} and \ce{Ba^{2+}}, on three natural clays: Wyoming montmorillonite (SWy-1), Georgia kaolinite (KGa-1), and Montana illite (IMt-1). The choice of these ions is driven by their properties, with sodium being among the most common monovalent cations in nature, and barium being a divalent heavy metal nuclear fission product, a component of nuclear waste used as a laboratory analogue to the more dangerous radioactive radium.\cite{Atun2003adsorption, Klinkenberg, Zhang2001adsorption} 


\subsection {Why clay mineral structures are so interesting?}

Typically, the term \emph{clay mineral} refers to hydrous phyllosilicates, further classified based on the chemophysical characteristics (see Table \ref{tab:Clay_summary}) arising from their structures. Since this work focuses on developing representative atomistic models of clay minerals, we must first review their structures and associated nomenclature.
 
\begin{figure}
    \centering
    \includegraphics[width=1\linewidth]{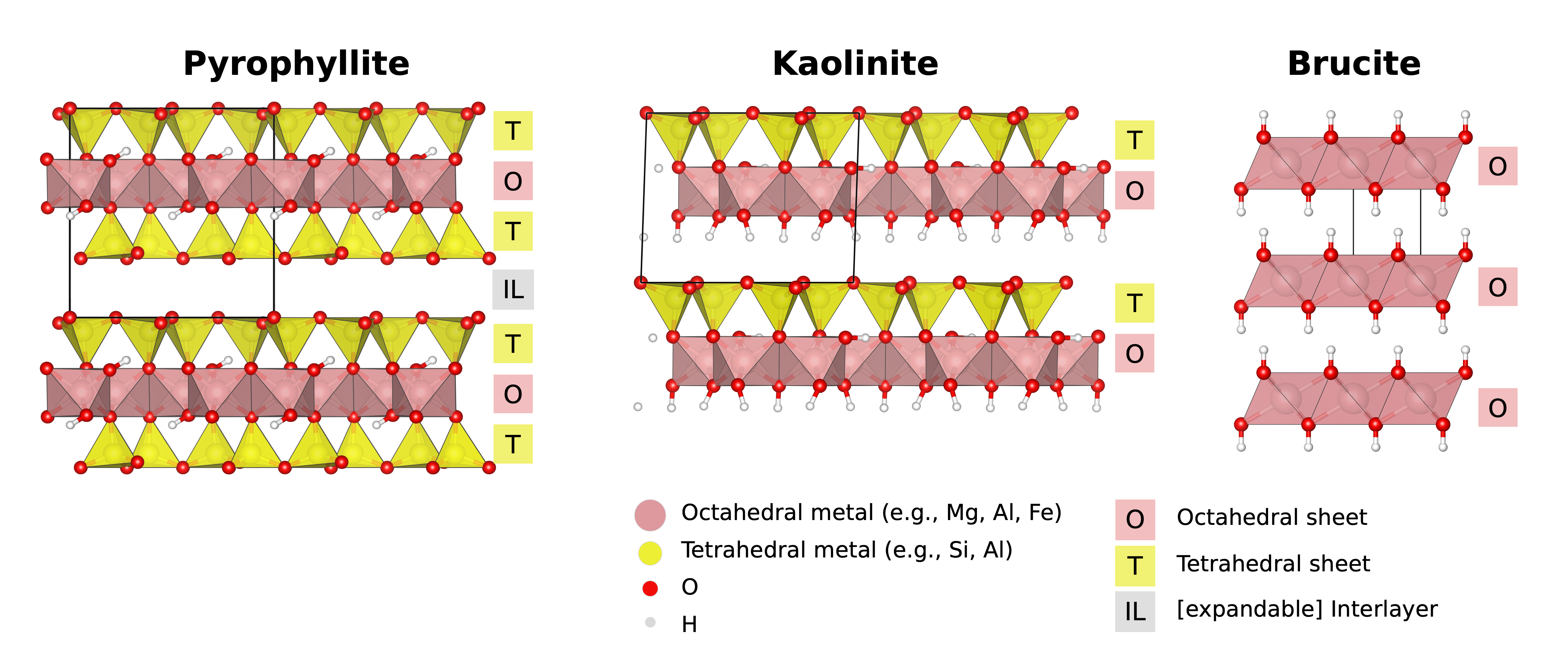}
    \caption{View onto three example representative clays: 2:1 pyrophyllite, 1:1 kaolinite and anionic clay brucite. The unit cell is shown in black, \ac{T} sheet in yellow, \ac{O} sheet in pink, and \ac{IL} space marked in blue.}
    \label{fig:bru-kao-pyr}
\end{figure}

At an atomic level, clay minerals are layered sheet silicate minerals. Each layer is composed of a stack of \ac{T} sheets bridging to \ac{O} sheets.
\ac{T} sheets are made up of a hexagonal network of \ce{SiO4}-tetrahedra, which are connected via three of their four oxygen atoms, the remaining apical oxygen links to the \ac{O} sheet.
\ac{O} sheets feature octahedra of six-fold divalent or trivalent metal cations (M\textsuperscript{2+/3+}), e.g., \ce{Al^{3+}}, \ce{Fe^3+}, \ce{Fe^2+} or \ce{Mg^2+}. 
The \ac{O} sheet metal ion valency is \ce{M^3+} in \ac{diO} and \ce{M^2+} in \ac{triO} clays.
In \ac{diO} clays, only two out of three sites within the \ac{O} sheet are occupied. Depending on the vacancy position relative to the hydroxyl groups, there exist \textit{cis}- and \textit{trans}-vacant varieties.
\Ac{T} and \ac{O} sheets are connected via the apical oxygen atoms of the \ac{T} sheet bridging to the octahedral metal cations. If the \ac{O} sheet is only bound to one \ac{T} sheet, the clay has a 1:1  structure (TO), whereas if the \ac{O} sheet is sandwiched between two \ac{T} sheets, the clay is of 2:1 type (TOT). 
Clay mineral geometries are usually defined via \acp{UC}, where one \ac{UC} represents the smallest periodically extendable unit of a clay layer. Figure \ref{fig:bru-kao-pyr} present schematics of possible \acp{UC}, and Table \ref{tab:Clay_summary} details example \ac{UC} compositions.

In a neutral clay, all \ac{T} sites are occupied with \ce{Si^{4+}}, and \ac{O} sites contain \ce{M^{2+}} metal cations for \ac{triO}, or \ce{M^{3+}} for \ac{diO} clays (e.g., pyriophillyte). Permanent negative layer charges arise from isomorphic substitutions (\ac{T}: \ce{Si^4+}~$\rightarrow$~\ce{M^3+}, \ac{diO}: \ce{M^3+}~$\rightarrow$~\ce{M^2+}, \ac{triO}: \ce{M^2+}~$\rightarrow$~\ce{M^1+}) or vacancies. This charge is compensated by cations in the interlayer space.

Systems with low to medium layer charge (\SIrange{-0.2}{-0.6}{\elementarycharge\per\uc}) are swelling clays (e.g., smectite). Their interlayer is populated by hydrated cations which are exchangeable for other ions or molecules. Swelling occurs through an exchange of interlayer species or varying hydration of the interlayer, leading to an expansion or contraction of the \textit{d}-space (defined as the sum of the clay layer thickness and interlayer space).
Initial expansion proceeds in discrete steps that correspond to the formation of up to four water layers (\textit{d}-space from 10 to 22 Å), and is referred to as crystalline swelling.
Beyond this distance (\textit{d}-space $>$ 22 Å), adjacent clay layers are no longer coordinated. Any further continuous increase in separation is called osmotic swelling.

At high layer charges, between \SIrange{-0.6}{-0.9}{\elementarycharge\per\uc}, the electrostatic attraction between clay and interlayer cations is too strong to permit an expansion beyond one to two water layers (e.g., vermiculite and true mica). 
As layer charges increase even further, reaching up to \SI{-1.0}{\elementarycharge\per\uc} for brittle mica, clay-cation attractive forces become too strong for any swelling to occur.

Unlike 2:1 clays formed by TOT layers, i.e., always exposing \ac{T} sheets to the interlayer, the 1:1 clays are made of TO sheets and expose two different types of surfaces: a hydrophobic siloxane and a hydrophilic hydroxide (see kaolinite on Figure \ref{fig:bru-kao-pyr}).
Typically, 1:1 clays do not feature large amounts of isomorphic substitutions and, therefore, do not have a permanent charge.

Another class of minerals we must mention is \acp{LDH} (see brucite on Figure \ref{fig:bru-kao-pyr}). Although technically not clay minerals, \acp{LDH} share many of their properties including a layered structure, isomorphic substitutions producing variable layer charge density, ion-exchange properties, and the ability to swell and interchalate. \acp{LDH} only feature octahedral sheets, where substitutions produce a positive charge. For example, in a hydrotalcite \ce{Mg^{2+}} may be replaced by \ce{Al^{3+}}, giving rise to a positive charge that is often counter-balanced by a \ce{CO3^{2-}} anion in the interlayer. For this reason, \acp{LDH} are sometimes referred to as \emph{anionic clays}. While present in nature, LDHs are also commonly synthesised in a laboratory setting. Since control over their composition is straightforward, it is commonplace to name them by only stating the ratio of \ce{M^{2+}} to \ce{M^{3+}}. For instance, an hydrotalcite with a \ce{[Mg6Al2(OH)16](CO3)(H2O)4} composition can be simply named 3:1 MgAl-\ac{LDH}.

\begin{table}[H]
\centering
\caption{Classification of clay minerals, including phyllosilicates and layered double hydroxides. Partially adapted from \citeauthor{Martin1991report}.\cite{Martin1991report} The example clay species are all planar unless otherwise specified, the example species that can be constructed with \textit{ClayCode} are highlighted in \textbf{bold}.}
\label{tab:Clay_summary}
\resizebox{\textwidth}{!}{%

\begin{tabular}{llll}
\hline
\begin{tabular}[c]{@{}l@{}}Interlayer Species\\ (Interlayer Charge, \\ ~~q,\si{\elementarycharge\per\uc})\end{tabular} &
  Group &
  \begin{tabular}[c]{@{}l@{}}Octahedral Type \\ Example UC Composition\end{tabular} &
  Example Species \\ \hline
\multicolumn{4}{c}{1:1 clay} \\ \hline
\multirow{5}{*}{\begin{tabular}[c]{@{}l@{}}none/ water\\ (q $\sim$ 0)\end{tabular}} &
  \multirow{3}{*}{serpentinite} &
  \multirow{3}{*}{\begin{tabular}[c]{@{}l@{}}triO\\ \ce{[Mg3][Si2]O5(OH)4}\end{tabular}} &
  \textbf{lizardite, berthierine, amesite, cronstedtite, nepouite} \\
 &
   &
   &
  \begin{tabular}[c]{@{}l@{}}Modulated sheets: \\ - Td-strips: antigorite, bemenitite\\ - Td-islands: greenalite, caryopilite, pyrosmalite\end{tabular} \\
 &
   &
   &
  Rolled/spheroidal: chrysotile, pecoraite \\ \cline{2-4} 
 &
  \multirow{2}{*}{kaolinite} &
  \multirow{2}{*}{\begin{tabular}[c]{@{}l@{}}diO\\ \ce{[Al2][Si2]O5(OH)4}\end{tabular}} &
  \textbf{kaolinite, dickite, nacrite, halloysite} \\
 &
   &
   &
  Rolled/spheroidal: halloysite \\ \hline
\multicolumn{4}{c}{2:1 clay} \\ \hline
\multirow{3}{*}{\begin{tabular}[c]{@{}l@{}}none \\ (q $\sim$ 0)\end{tabular}} &
  \multirow{2}{*}{talc} &
  \multirow{2}{*}{\begin{tabular}[c]{@{}l@{}}triO\\ \ce{[Mg3][Si4]O10(OH)2}\end{tabular}} &
  \textbf{talc, willemseite, kerolite, pimelite} \\
 &
   &
   &
  \begin{tabular}[c]{@{}l@{}}Modulated sheets:\\ - Td-strips: minnesotaite\end{tabular} \\ \cline{2-4} 
 &
  pyrophyllite &
  \begin{tabular}[c]{@{}l@{}}diO\\ \ce{[Al2][Si4]O10(OH)2}\end{tabular} &
  \textbf{pyrophyllite, ferripyrophyllite} \\ \hline
\multirow{4}{*}{\begin{tabular}[c]{@{}l@{}}hydrated \\ exchangeable cations, M$^+$\\ (q $\sim$ 0.2-0.6)\end{tabular}} &
  \multirow{4}{*}{smectite} &
  \multirow{2}{*}{\begin{tabular}[c]{@{}l@{}}triO\\ \ce{(M$^+ _y \cdot $ n H2O) [Mg$_{3-y}$Li$_y$] [Si4] O10(OH)2}\end{tabular}} &
  \textbf{saponite, hectorite, sauconite, stevensite, swinefordite} \\
 &
   &
   &
  \begin{tabular}[c]{@{}l@{}}Modulated sheets:\\ - Oct-strips: sepiolite, loughlinite\end{tabular} \\ \cline{3-4} 
 &
   &
  \multirow{2}{*}{\begin{tabular}[c]{@{}l@{}}diO \\ \ce{(M$^+ _y \cdot $ n H2O) [Al$_{2-y}$Mg$_y$] [Si4] O10(OH)2}\end{tabular}} &
  \textbf{montmorillonite, beidellite, nontronite, volkonskoite} \\
 &
   &
   &
  \begin{tabular}[c]{@{}l@{}}Modulated sheets:\\ - Oct-strips: palygorskite, yofortierite, windhoekite\end{tabular} \\ \hline
\multirow{2}{*}{\begin{tabular}[c]{@{}l@{}}hydrated \\ exchangeable cations\\ (q $\sim$ 0.6-0.9)\end{tabular}} &
  \multirow{2}{*}{vermiculite} &
  \begin{tabular}[c]{@{}l@{}}triO\\ (see smectite)\end{tabular} &
  \textbf{trioctahedral vermiculite} \\ \cline{3-4} 
 &
   &
  \begin{tabular}[c]{@{}l@{}}diO\\ (see smectite)\end{tabular} &
  \textbf{dioctahedral vermiculite} \\ \hline
\multirow{3}{*}{\begin{tabular}[c]{@{}l@{}}non-hydrated\\ monovalent cations \\ (q $\sim$ 0.6-1.0)\end{tabular}} &
  \multirow{3}{*}{true (flexible) mica} &
  \multirow{2}{*}{\begin{tabular}[c]{@{}l@{}}triO\\ \ce{K [Mg3](AlSi3)O10(OH)2}\end{tabular}} &
  \textbf{biotite, phlogopite, lepidolite} \\
 &
   &
   &
  \begin{tabular}[c]{@{}l@{}}Modulated sheets:\\ - Td-strips: ganophyllite, eggletonite, tamaite, bannisterite\\ - Td-islands: zussmanite, coombsite\end{tabular} \\ \cline{3-4} 
 &
   &
  \begin{tabular}[c]{@{}l@{}}diO\\ \ce{K [Al2][AlSi3]O10(OH)2}\end{tabular} &
  \textbf{muscovite, illite, glauconite, celadonite, paragonite} \\ \hline
\multirow{2}{*}{\begin{tabular}[c]{@{}l@{}}non-hydrated\\ divalent cations \\ (q $\sim$ 1.8-2.0)\end{tabular}} &
  \multirow{2}{*}{brittle mica} &
  \begin{tabular}[c]{@{}l@{}}triO\\ \ce{Ca [Mg2Al](AlSi3)O10(OH)2}\end{tabular} &
  \textbf{clintonite, kinoshitalite, bityite, ananditc} \\ \cline{3-4} 
 &
   &
  \begin{tabular}[c]{@{}l@{}}diO\\ \ce{Ca [Al2][Al2Si2]O10(OH)2}\end{tabular} &
  \textbf{margarite, cherbykhite} \\ \hline
\multicolumn{4}{c}{mixed layer clays} \\ \hline
\multirow{2}{*}{\begin{tabular}[c]{@{}l@{}}hydroxide sheet \\ (q variable)\end{tabular}} &
  \multirow{2}{*}{chlorite} &
  \begin{tabular}[c]{@{}l@{}}triO\\ \ce{Mg3(OH)6 [Mg2Al] [AlSi3] O10)(OH)2 }\end{tabular} &
  clinochlore, chamosite, pennantite, nimite, baileychlore \\ \cline{3-4} 
 &
   &
  \begin{tabular}[c]{@{}l@{}}diO\\ \ce{Al2(OH)6 [Al2][Si3Al]O10(OH)2 }\end{tabular} &
  donbassite \\ \hline
\multirow{2}{*}{\begin{tabular}[c]{@{}l@{}}regularly interstratified\\ (q variable)\end{tabular}} &
  \begin{tabular}[c]{@{}l@{}}mix of two types\\ (see species example)\end{tabular} &
  triO &
  \begin{tabular}[c]{@{}l@{}}corrensite (chlorite+smectite/vermiculite), \\ aliettite (talc+saponite), hydrobiotite (biotite+vermiculite), \\kulkeite (mica+vermiculite)\end{tabular} \\ \cline{3-4} 
 &
   &
  diO &
  \begin{tabular}[c]{@{}l@{}} rectorite (muscovite/illite+montmorillonite),\\ tosudite (chlorite+smectite) \end{tabular} \\ \hline

\multicolumn{4}{c}{anionic clays} \\ \hline

\multirow{2}{*}{\begin{tabular}[c]{@{}l@{}}none \\ (q $\sim$ 0)\end{tabular}} &  \multirow{2}{*}{brucite} &
  {\begin{tabular}[c]{@{}l@{}} triO\\ \ce{Mg(OH)2}\end{tabular}} &  \textbf{brucite, amakinite} \\ \cline{2-4} 
 &   gibbsite &  \begin{tabular}[c]{@{}l@{}} diO\\ \ce{Al(OH)3}\end{tabular} &  \textbf{gibbsite, bayerite} \\ \hline
  
  \begin{tabular}[c]{@{}l@{}}water + anions,\\ e.g., \ce{CO3^{2-}}, \ce{NO3-}, \ce{Cl-} \\(q $\sim$ -0.2 -- -1.0)\end{tabular} & hydrotalcite &
  \begin{tabular}[c]{@{}l@{}} triO\\ \ce{[Mg6Al2(OH)16](CO3)(H2O)4}\end{tabular} & \textbf{quintinite, pyroaurite, iowaite,  fougèrite} \\ \hline
    
\end{tabular}
}
\end{table}

Table \ref{tab:Clay_summary} summarises clay classification, listing the core nine groups of silicate clay minerals as well as anionic clays, alongside example mineral species. 
It can be seen that not all clays are planar: some involve modulated sheets connecting via either \ac{O} or \ac{T} sheets, and some are rolled. While fascinating, these cases are relatively uncommon and thus are not the focus of the current discussion, and are not currently implemented in \textit{ClayCode}. Most importantly, with this summary, we wish to emphasise the vast variety of clay species originating from the small changes in the clay structure. 

For instance, montmorillonite is among the most widely studied clays, thanks to its powerful adsorption abilities and common presence in soils around the world. Well-characterised samples can be purchased from the Clay Minerals Society\cite{Clayminsoc} for experimental work. Within these samples there exist a few sub-types:
\begin{enumerate}
\item Texas montmorillonite (STx-1):\\
    $\mathrm{
    \big[[Si_{8.00}]
    [Al_{2.41} Fe(III)_{0.09} Mn_{<0.01} Mg_{0.71} Ti_{0.03}]O_{20}(OH)_{4}\big]
    ^{-0.68}}$
\item Wyoming montmorillonite (SWy-1, SWy-2, SWy-3):\\
    $\mathrm{
    \big[[Si_{7.98}Al_{0.02}]
    [Al_{3.01} Fe(III)_{0.41}Mn_{0.01}Mg_{0.54} Ti_{0.02}]O_{20}(OH)_{4}\big]^{-0.55}}$
\item Otay montmorillonite (SCa-2):\\
    $\mathrm{
    \big[[Si_{7.81}Al_{0.19}]
    [Al_{2.55} Fe(III)_{0.12}Mn_{<0.01}Mg_{1.31} Ti_{0.02}]_{20}(OH)_{4}\big]^{-1.48}}$
\end{enumerate}
Yet, this diversity has not been reflected in the simulations up to now. Indeed, montmorillonite is commonly simulated using one of the following four idealised structure models:
\begin{enumerate}
\item $\mathrm{
    \big[[Si_{8}][Al_{3}Mg_{1}]O_{20}(OH)_{4} \big]^{-1.00}}$ ~\cite{Underwood2016, Underwood2015, Zhang2014, Zhang2016surface} 
\item  $\mathrm{
    \big[[Si_{7.75}Al_{0.25}][Al_{3.25}Mg_{0.75}]O_{20}(OH)_{4} \big]^{-1.00}}$ ~\cite{Zheng2010} 
\item $\mathrm{
    \big[[Si_{7.75}Al_{0.25}][Al_{3.5}Mg_{0.5}]O_{20}(OH)_{4}\big]^{-0.75}}$  ~\cite{Chang1998, Smith1998, Chang1995, Boek1995, Rahromostaqim2018, Ngouana2014structural}
\item $\mathrm{
    \big[[Si_{8}][Al_{3.25}Mg_{0.75}]O_{20}(OH)_{4}\big]^{-0.75}}$ \cite{Zhou2014, Teich-McGoldrick2015, Rotenberg2010, Rotenberg2009, Zhang2016surface} 
\end{enumerate}


\subsection {What tools are available and why do we need another?}

Determining the structure and parameters of a clay model is a complex and laborious process. This task has to be accomplished either manually or with software tools of limited applicability, which explains why to date clay models used in simulations are greatly simplified.
Historically, the most common tool used to prepare MD simulations of clay systems has been the commercial software Materials Studio.\cite{Materials-studio}
This software enables the user to create custom material structures via a graphical user interface and offers tools to help assign force field parameters. While useful, Materials Studio's applicability is limited by the ability of the user to accurately draw out the desired clay or load in the desired CIF files, reassigning the atoms by hand. Furthermore, its force field assignment needs to be manually checked and validated, which is itself a laborious process.

In the last five years, two freely available tools have also become available. The \textit{atom}\cite{Holmboe2019} library leverages on MATLAB to enable setting up a wide range of periodic inorganic structures.
While flexible, it requires programming knowledge, and cannot be used as a standalone application. 
The more recent CHARMM-GUI \textit{Nanomaterial Modeler}\cite{Choi2021}, while more accessible than \textit{atom}, can only handle four simplified clay species (montmorillonite, kaolinite, pyrophyllite, and muscovite).

The \textit{ClayCode} Python package automates the setup of clay models for classical \ac{MD} simulations, offering tools to design realistic clay models and to set up custom simulation workflows. 
\textit{ClayCode} has been designed to be user-friendly: while its behaviour is fully customizable, its default parameters enable simulation of most common systems with minimal user intervention. 
In its current embodiment, \textit{ClayCode} enables modelling planar hydrated clay systems ready for simulation with GROMACS.\cite{Berendsen1995Gromacs} Clay sheets are assembled from \acp{UC} modelled with the mainly non-bonded ClayFF force field parameters.\cite{Cygan2004,Cygan2021} By default, water and ions are described according to the \ac{SPC} model\cite{Berendsen1987SPC} and ion parameters by \citeauthor{Smith2023ConsequencesIons}\cite{Smith2023ConsequencesIons}; however, the user can choose alternative models. Table \ref{tab:tools} provides a side-by-side comparison of CHARMM-GUI \textit{Nanomaterial Modeler}, \textit{atom}, and our \textit{ClayCode}.

\begin{table}[H]
\centering
\caption{Comparison of CHARMM-GUI \textit{Nanomaterial Modeler}, \textit{atom} and \textit{ClayCode}.\\
(* indicates that compatibility is available for ClayFF only.)}
\label{tab:tools}
\resizebox{\textwidth}{!}{

\begin{tabular}{c|ccc}
              & \emph{\textbf{Nanomaterial Modeler}} & \emph{\textbf{atom}}                                              & \emph{\textbf{ClayCode}} \\ \hline
Language      & Tcl/Tk                                   & MATLAB                                                     & Python            \\ \hline
Force Field   & INTERFACE                                & \begin{tabular}[c]{@{}c@{}}ClayFF\\ INTERFACE \end{tabular} & ClayFF            \\ \hline
\multirow{6}{*}{\begin{tabular}[c]{@{}c@{}}Simulation Engine\\ Compatability\end{tabular}} &
  \multirow{6}{*}{\begin{tabular}[c]{@{}c@{}}AMBER\\ CHARMM \\ GENESIS\\ LAMMPS \\ NAMD\\ OpenMM \end{tabular}} &
  \multirow{6}{*}{\begin{tabular}[c]{@{}c@{}}GROMACS \\ NAMD*\\ LAMMPS*\\ RASPA2*\end{tabular}} &
  \multirow{6}{*}{GROMACS} \\
              &                                          &                                                            &                   \\
              &                                          &                                                            &                   \\
              &                                          &                                                            &                   \\
              &                                          &                                                            &                   \\ 
              &                                          &                                                            &                   \\ 
              &                                          &                                                            &                   \\ \hline
\multirow{3}{*}{UC database provided} &
    \begin{tabular}[c]{@{}c@{}}\ce{[Al4][Si_4]O10(OH)8}\\ \ac{diO} 1:1 kaolinite\end{tabular} &
    \multirow{3}{*}{no} &
    \multirow{3}{*}{list in Table \ref{tab:uc_types}} \\
              &                                          &                                                            &                   \\ 
 &
  \begin{tabular}[c]{@{}c@{}}\ce{[[Al_{4-$m$}Mg_{$m$}][Si_{8-$n$}Al_{$n$}]O20(OH)4]^{-($m$+$n$)}}\\ \emph{cis}-vacant \ac{diO} 2:1 \\ pyrophyllite (n=0, m=0), \\ montmorillonite (m$\neq$0, n=0)\\ muscovite (m$\neq$0, n$\neq$0)\end{tabular} &
   &
   \\ \hline
User UC input & not possible                             & yes                                                        & yes               \\ \hline
\end{tabular}
}
\end{table}


\section{The \textit{ClayCode} Workflow}

\textit{ClayCode} runs in the terminal of Unix-based operating systems. It automatically builds and prepares for the simulations periodic atomistic models of hydrated planar clay sheets using an internal customisable database of \acp{UC}.
To this end, the \emph{ClayCode} workflow is subdivided into a set of steps, handled by independent modules, explicitly designed to facilitate the addition of new functionalities.
Currently, the available modules are \texttt{data} (with an internal \ac{UC} and force field database where custom \acp{UC} and force fields can be added), \texttt{builder} (to assemble the clay system), and \texttt{siminp} (to generate associated GROMACS input files).
The user can then analyse the produced GROMACS trajectories with typical MD analysis tools, e.g., GROMACS own ones,\cite{Berendsen1995Gromacs} MDAnalysis \cite{Michaud-Agrawal2011}, VMD\cite{Humphrey1996vmd}, or our own DynDen\cite{Degiacomi2021dynden}. The user interacts with each \textit{ClayCode} module via plain text files in YAML format. 
Hereafter, we describe the main functionalities of each module. A full list of all available keywords, along with their default values, can be found in the online documentation at \href{https://claycode.readthedocs.io/en/latest/}{claycode.readthedocs.io}.

\subsection {Handling unit cells with the \texttt{data} module}

Clay sheets are made from an internal \ac{UC} database containing a selection of different unit cell types that have been either constructed from the \ac{AMCSD} crystallographic data or taken from our previous works (see summary in Table \ref{tab:uc_types}). 
The availability of pre-assigned \acp{UC} allows constructing a large number of clays without the need to dwell into crystallographic data and assigning ClayFF atom types to the \acp{UC}. Nevertheless, the user can also expand the database as needed.

Adding new \acp{UC}, based upon already assigned unsubstituted ones, can be done via the \emph{data} module. In this step, \acp{UC} containing all possible substitutions are produced. This operation can be repeated iteratively, to produce \acp{UC} featuring up to three substitutions. For all generated \acp{UC}, as default the maximum accepted charge limit arising from isomorphic substitutions is set to \SI{-2.0}{\elementarycharge\per\uc}. Any \ac{UC} with substitution combinations resulting in higher charges will not be added to the database.

  \begin{table}[H]
      \caption{Currently available unit cell types in \textit{ClayCode} \ac{UC} database. \ac{UC} constructed from structures in the \ac{AMCSD}\cite{Downs2003}, or manually curated by us for this work or in the earlier studies.\cite{Underwood2015, Underwood2016, Erastova2017understanding, Erastova2017, Tian2018understanding, Zhao2024revealing}} 
      \label{tab:uc_types}
          \begin{tabularx}{\linewidth}{ccccc}
              \toprule
              \textbf{UC type} & \textbf{UC Description}        & \textbf{AMCSD code} & \textbf{example} & \textbf{Ref.} \\
              \midrule 
              TD21                  & \textit{trans}-\ac{diO} 2:1 & 0007180    &  nontronite & \cite{Dainyak2006} \\ 
              CD21                  & \textit{cis}-\ac{diO} 2:1   & 0002868    &  montmorillonite & \cite{Viani2002, Underwood2015}  \\ 
              TD11                  & \textit{trans}-\ac{diO} 1:1  &   -       & dickite  &         \\
              CD11                  & \textit{cis}-\ac{diO} 1:1    &  -        & kaolinite  & \cite{Underwood2016, Tian2018understanding, Zhao2024revealing}        \\
              T21                   & \ac{triO} 2:1         & 0015819          & saponite  & \cite{Breu2003}    \\
              T11                   & \ac{triO} 2:1         &    -             & lizardite  &         \\
              LDH21                 & 2:1 LDH               &  -               & quintinite  &         \\
              LDH31                 & 3:1 LDH               & 0007912          & hydrotalcite   & \cite{Catti1995, Erastova2017understanding, Erastova2017}     \\
              \bottomrule
          \end{tabularx}
\end{table}


\subsection{Setting up clay models with the \texttt{builder} module}

\begin{figure}[H]
\includegraphics[width=0.5\textwidth]{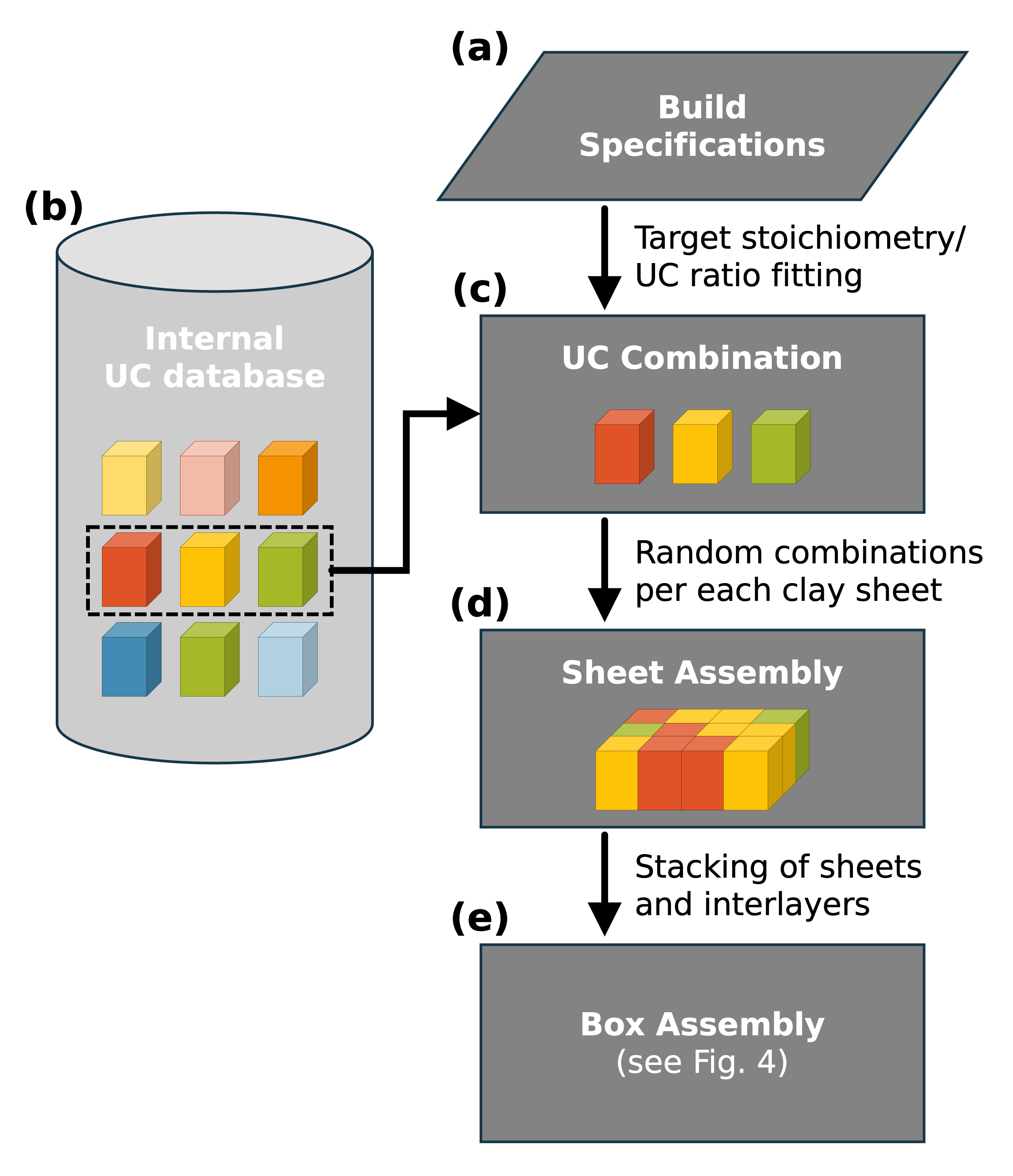}
\caption{Illustration of sheet assembly workflow. Every clay system is made of individual clay sheets, each featuring a random arrangement of known \ac{UC} in proportions to match user-defined specifications.}
\label{fig:assembly}
\end{figure}

A hydrated clay system is assembled according to a four-step pipeline, illustrated in Figure \ref{fig:assembly}, ultimately yielding a topology and a coordinates file ready for simulation with GROMACS. Hereafter, we describe each step.

\noindent \textbf{ I. Parsing and checking user-provided input data}

\noindent The minimal input for model building requires a system name, the \ac{UC} type, and either a specification of the average \ac{UC} stoichiometry, or ratios of \ac{UC} indices from which the sheets should be assembled (Figure \ref{fig:assembly} a).


\noindent \textbf{ II. Matching target \ac{UC} composition}

\noindent Using the \ac{UC} database (Figure \ref{fig:assembly} b), \textit{ClayCode} will either find the combination of \acp{UC} that provides the closest match to a user-defined target stoichiometry, or compute the numbers of each \ac{UC} that correspond to given \ac{UC} ratios (Figure \ref{fig:assembly} c).
\emph{ClayCode} only adds substitutions of elements with a minimum occupancy of \SI{0.05} {atoms\per\uc}. 
In the case of substitution not available in the \ac{UC} database, those under threshold are removed and the remaining sheet occupancies are adjusted to match the average target charges. For the ones above the threshold, the user is prompted to specify the oxidation state of the incompatible atom type which is then also removed from the target \ac{UC} charge.


\noindent \textbf{III. System assembly}

\begin{figure}
    \centering
    \includegraphics[width=1\linewidth]{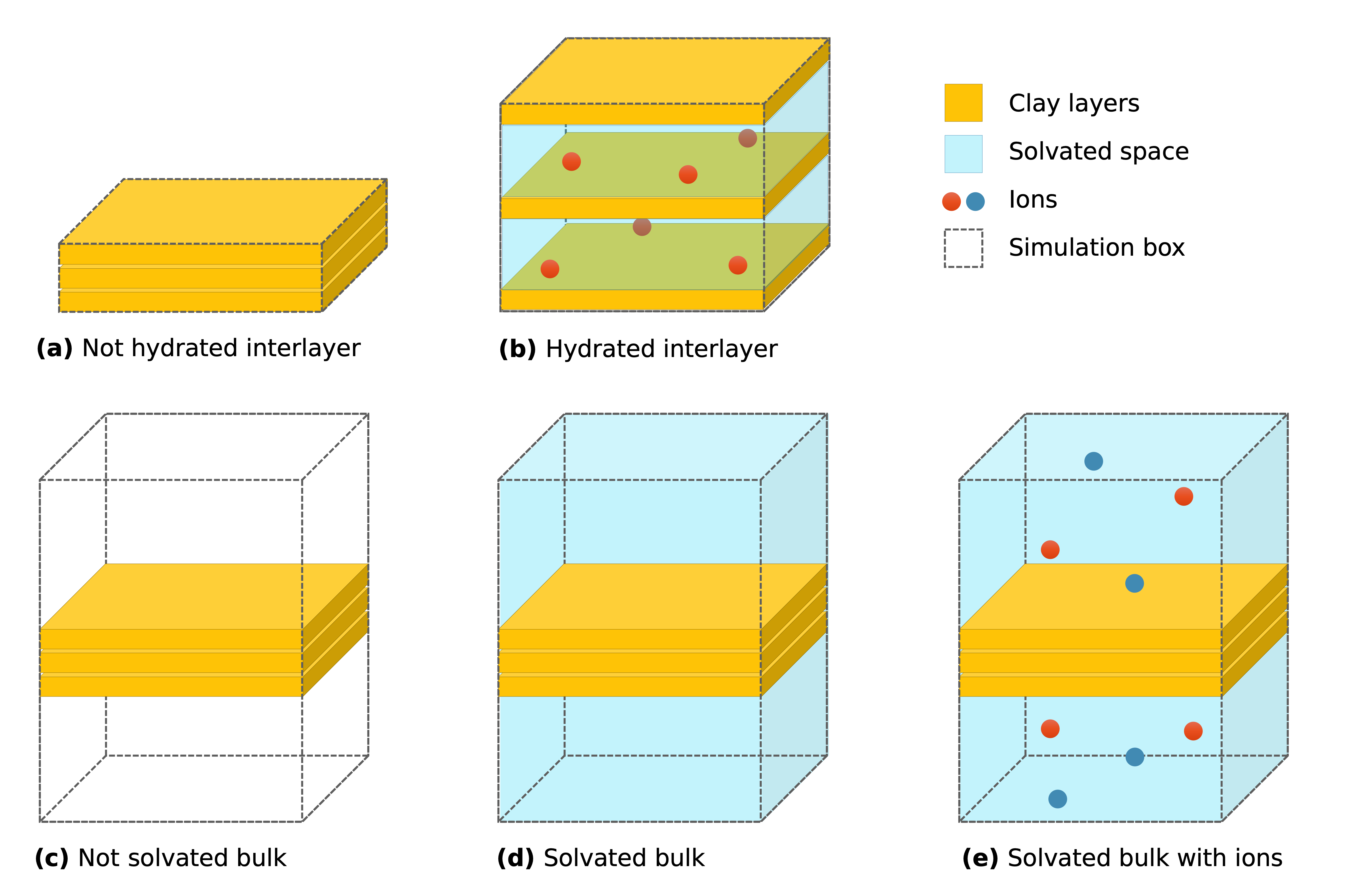}
    \caption{Illustration of clay models with different sheet stackings. 
        Stacked clay sheets can either be without interlayer water (a) or with a hydrated interlayer (b). The different bulk space setup options also offer setup of a non-solvated (c), solvated with water only (d) or with added ions (e) systems.}
    \label{fig:stackings}
\end{figure}

\noindent Sheets are assembled from a randomised sequence of the selected \acp{UC} (Figure \ref{fig:assembly} d). This randomisation is important to represent a realistic clay structure, where substitutions are randomly distributed within the sheets.
Combinations of \acp{UC} yield clay sheets with substitutions that obey the Loewenstein rule.\cite{Lowenstein1954} 
Then, the interlayer is generated using GROMACS \texttt{solvate} and \texttt{genion} tools.
The sheets and interlayers are finally stacked, the layer charges resulting from isomorphic substitutions are compensated by counterions, as selected by the user; and, if specified, solvent and bulk ions are added to an extended simulation box bulk space (Figure \ref{fig:assembly} e). Currently supported box arrangements are illustrated in Figure \ref{fig:stackings}.

It has to be noted that the accuracy of a clay model is limited by the number of \acp{UC} within each sheet. In particular, to achieve a close match to a target composition with minor quantities of substitutions, a minimum sheet size is required. By default, \textit{ClayCode} accepts absolute deviation of occupancies from the target structure below \SI{0.025}{atoms\per\uc}.

\noindent \textbf{ IV. Energy minimisation}

\noindent The final step is a GROMACS energy minimisation run. 
All resulting output (topology, coordinates, and a log file reporting on \emph{ClayCode} operations) is then placed inside an output folder with the system's name.


\subsection{Writing simulation run scripts with \textit{siminp} module}

The \textit{siminp} module produces all the files required for custom molecular dynamics pipelines. To this end, the user specifies a series of run types and the GROMACS version that will be used to run the simulations. By default, \textit{ClayCode} will produce an output to be run with the GROMACS version which is installed locally.
This input is then processed by \textit{ClayCode} and a directory tree with all required input files (coordinates, topology, force field and run parameters) and a run script is constructed. 
The run parameter MDP files are assembled based on the allowed parameters for the selected GROMACS version. Then, if specified, GROMACS default parameters are overwritten with run-specific user-defined MDP options. 

One of the features of \textit{siminp} is its ability to generate \textit{d}-space equilibration run input files, whereby water molecules are iteratively removed from a hydrated interlayer until the distance between clay sheets converges to a specified distance.


\section{Methods}

\subsection {Clay models}

KGa-1, SWy-1 and IMt-1 clay models were constructed using \textit{ClayCode} to match experimental compositions of Clay Minerals Society source clays Georgia kaolinite KGa-1, Wyoming montmorillonite SWy-1/SWy-2/SWy-3 and Montana illite IMt-1/IMt-2, respectively. 
Clay sheets were assembled from \acp{UC} available within the database: CD21 \acp{UC} for SWy-1 and IMt-1, and CD11 \acp{UC} for KGa-1. 
Furthermore, we also produced a simplified montmorillonite model (hereon SWy-simp.) matching that used in our previous work.\cite{Underwood2015, Underwood2016} To this end, clay sheets were assembled from a single $cis$-\ac{diO} D21 \ac{UC} with one \ac{O} substitution of Al\textsuperscript{3+} for Mg\textsuperscript{2+} per \ac{UC}. This was done by selecting \textit{ClayCode}'s \ac{UC} ratio method rather than giving a target composition. 

Table~\ref{tab:all-comp} summarises the experimental and modelled clays' stochiometries. 
KGa-1 and IMt-1 have non-exchangeable interlayer ions, therefore, those are matched to the reference structures.
Montmorillonite is a swelling clay with exchangeable interlayer ions. Typically, in the laboratory, before starting adsorption studies, swelling clays undergo homoionisation. To replicate this, we have set the interlayer ions to be \ce{Ca^{2+}}. This interlayer space in swelling clays is variable and is influenced by the ionic composition and hydration levels. For a Ca-rich hydrated montmorillonite the \textit{d}-space is generally around \SI{1.5}{\nm}, which is what we have also set here.
The YAML files and the experimental composition file used to set up these systems are all included as Supporting Information. We note that, as the \acp{UC} arrangement is randomly generated, each new run of \textit{ClayCode} will produce a slightly different model.

\begin{table}[H]
\caption{Average \ac{UC} stoichiometries of simplified (SWy-simp.) and realistic SWy-1, KGa-1 and IMt-1 models and their corresponding experimental reference (ref.) data from Clay Mineral Society.\cite{Clayminsoc} The interlayer ions are given in round brackets.}
\label{tab:all-comp}
\resizebox{\textwidth}{!}{%
\begin{tabular}{ll}
    \toprule
        Clay system & Average UC stoichiometry \\
    \midrule
        SWy-simp. model & \ce{(Ca_{0.48}Na_{0.04})[[Si_8][Al_3 Mg]O_{20}(OH)4]^{-1.00}} \\
        SWy-1 model &   \ce{(Ca_{0.25}Na_{0.04})[[Si7_.97Al0_.03][Al3_.02Fe^{III}0_.43Mg0_.54]O_{20}(OH)4]^{-0.54}}\\
        SWy-1 ref.  &  \ce{(Ca_{0.12}Na_{0.32}K_{0.05})[[Si_{7.98}Al_{0.02}][Al_{3.01}Fe^{III}_{0.41}Mn_{0.01}Mg_{0.54}Ti_{0.02}]O_{20}(OH)4]^{-0.53}} \\ 
        \hline
        KGa-1 model & \ce{(Mg_{0.02}Ca_{0.02}Na_{0.05}K_{0.04})[[Si_{3.83}Al_{0.17}][Al_{3.97}Fe^{III}_{0.03}]O_{10}(OH)8]^{-0.17}} \\
        KGa-1 ref.  & \ce{(Mg_{0.02}Ca_{0.01}Na_{0.01}K_{0.01})[[Si_{3.83}Al_{0.17}][Al_{3.86}Fe^{III}_{0.02}Mn_{tr}Ti_{0.11}]O_{10}(OH)_8]^{-0.06}} \\
        \hline
        IMt-1 model & \ce{(Ca_{0.06}Mg_{0.09}K_{1.37})[[Si_{6.77}Al_{1.23}][Al_{2.74}Fe^{II}_{0.03}Fe^{III}_{0.83}Mg_{0.40}]O_{20}(OH)4]^{-1.67}} \\
        IMt-1 ref.  & \ce{ (Ca_{0.06}Mg_{0.09}K_{1.37})[[Si_{6.77}Al_{1.23}][Al_{2.69}Fe^{II}_{0.06}Fe^{III}_{0.76}Mn_{tr}Mg_{0.43}Ti_{0.06}]O_{20}(OH)4]^{-1.68}} \\
        \bottomrule
\end{tabular}%
}
\end{table}

The produced clay models were then solvated with \ac{SPC} water\cite{Berendsen1987SPC}, the excess charge was neutralised with \ce{Na+} and \ce{Ba^2+} ions.\cite{Smith2023ConsequencesIons} Then, the bulk ion concentrations were adjusted to equivalents of \SI{0.1}{\mole\per\liter} of \ce{Na+} and \ce{Ba^2+} using \ce{NaCl} and \ce{BaCl2} in equimolar quantities.
Average final simulation box dimensions, numbers of interlayer and bulk water, and numbers of inserted bulk ions are listed in Table~\ref{tab:ion_numbers}.


\subsection {Molecular dynamics simulations}

All of the simulations (energy minimisation, equilibration and production runs) were carried out with GROMACS 2022.3.\cite{Berendsen1995Gromacs}
Energy minimisation was performed at the end of the setup procedure with \textit{ClayCode}, using the default parameters set by \textit{ClayCode} -- steepest descent algorithm using as convergence criterion the maximum force on any one atom being less than \SI{500}{\kilo\joule \per\mol\per\nano\meter}.

This was followed by two short equilibration runs, each of \SI{0.5}{\nano\second} with a step size of \SI{0.5}{\femto\second}. The first one was performed in a Canonical (NVT) ensemble, with the Velocity-rescale thermostat set to \SI{300}{\kelvin} and the clay layers fixed along \emph{z}-direction. The second equilibration run was performed in the isothermal–isobaric (NPT) ensemble, adding Parinello-Rahman barostat set to \SI{1.0}{\bar} with semi-isotropic scaling to allow the decoupling of \emph{xy}-plane and \emph{z}-axis of the simulation box.

For the swelling clay models, i.e., SWy-simp. and SWy-1, this was followed by a sequence of \emph{d}-space equilibration runs, with the files generated with \emph{simpinp} module of \emph{ClayCode}. To this end, a number of interlayer water molecules (here, 50 molecules) is removed, then the system undergoes a short NPT equilibration that allows the \emph{z}-axis to contract, accounting for the new hydration level. The \emph{d}-space is then compared with a target value (here, \SI{1.5}{\nm}), and the dehydration step is repeated until an agreement is reached. For the systems in this work, it took six steps to obtain the desired \emph{d}-space. Each equilibration run is \SI{0.1}{\nano\second} long with \SI{1}{\femto\second} timestep, \SI{300}{\kelvin} and \SI{1.0}{\bar} controlled with Nose-Hoover and semi-isotropic Parrinello-Rahman algorithms, respectively.  

From there, the systems continue into production runs of \SI{70}{\nano\second} and a timestep of \SI{1.0}{\femto\second} in the isothermal–isobaric (NPT) ensemble with and Nose-Hoover thermostat at \SI{300}{\kelvin}  and semi-isotropic Parrinello-Rahman barostat at \SI{1.0}{\bar}. 

In all simulation runs, neighbour searching was performed every 10 steps, electrostatic and van der Waals interactions were computed using particle mesh Ewald algorithm with geometric combination rules, Verlet cutoff-scheme and \SI{1.4}{\nano\meter} cutoff distances. The LINCS algorithm was used for H-bond constraints. 

Production run trajectories were written at \SI{2}{\pico\second} intervals. The last \SI{50}{\nano\second} of each simulation were used for analyses, after the systems were assessed for convergence with DynDen\cite{Degiacomi2021dynden}.


\subsection {Analysis}

Linear number densities were calculated for each simulation using the GROMACS \texttt{gmx density} tool, with a window of 0.01 \AA. Densities were calculated independently for the clay (represented by all of its atoms), \ce{Ba^{2+}}, and \ce{Na^{+}} atoms (see Figure \ref{fig:S1}). We note there is some asymmetry observed in these plots. The asymmetry is an effect of the two clay layers not being exact mirror images of one other and ions positions being randomly initialised, where proximity to a strongly interacting site will result in the ion being trapped. This was previously observed and discussed when simulating a realistic clay system.\cite{Nuruzade2023organic} To accommodate for this phenomenon, observations for both sides of each clay can be aggregated.

All linear densities were analysed with an in-house Python code. For each simulation, we identified the position of clay boundaries along the $z$-axis (perpendicular to the clay surface) as the maxima in the first and last peak in its linear density (hereon $z_{upper}$ and $z_{lower}$, respectively). We used these boundaries to curate the measured ionic number densities. Specifically, we removed all ion density located between $z_{lower}$ and $z_{upper}$, i.e., ions in the interlayer, so only the distribution of ions in solution were considered. Each normalised density was then split in two parts, describing ionic distributions above and below the clay. Ionic densities above the clay were defined as those associated with a $z$-position greater than $z_{upper}$, and those below the clay as those with $z$-position lower than $z_{lower}$. The resulting two density distributions were shifted to bring their $z_{upper}$ and $z_{lower}$ position to the origin, respectively. Finally, the two distributions were overlaid by mirroring that of upper ions. SWy-simp., SWy-1, and IMt-1 systems feature clays with the TOT structure, i.e., exposing the same tetrahedral type of surface to the solvent, and were thus associated with comparable ion distributions. In these cases, the upper and lower ion densities were summed. In the case of KGa-1, the TO clay, these surfaces were presented independently, as KGa-1-T for the tetrahedral siloxane surface and KGa-1-O for the octahedral hydroxyl surface. Each obtained density distribution was finally independently normalised.

The locations of peaks in the number densities are associated with characteristic ion adsorption modes. By separately plotting the number densities of \ce{Ba^{2+}} and \ce{Na^{+}}, we identified common minima in distributions, that we took as criteria to define regions associated with each adsorption mode (see Figure \ref{fig:S2}). By integrating each region, we could quantify the percentage of ions adsorbed in each mode. For this quantification, for SWy-simp., SWy-1, and IMt-1 systems, upper and lower number densities were averaged. For KGa-1, we only considered the distribution of ions on the tetrahedral side.

\subsection{Visualisation}

Plots were produced with in-house software using the matplotlib\cite{Hunter2007} Python package. 
Renderings of clay \ac{UC} structures are made with VESTA.\cite{Momma2008vesta}
Simulation renderings were produced with VMD\cite{Humphrey1996vmd} with the colours as follows, unless stated otherwise. Clay layers are shown with balls and sticks representations, atom colours are: Al - pink, Mg - cyan, both \ce{Fe^{II}} and \ce{Fe^{II}} - green, O - red, H - white. Ionic species are shown as van der Waals spheres of different colours: \ce{Ba^{2+}} - orange, \ce{Na^+} - blue, \ce{Cl^-} - green, \ce{Mg^{2+}} - grey, \ce{Ca^{2+}} - silver, \ce{K^+} - black. Bulk water molecules are not shown for clarity, individual molecules of water are shown in balls and sticks representation with O - red and H - white.


\section{Results and Discussion}

\subsection {Construction of accurate clay models}

We used \textit{ClayCode} to construct clay models with compositions closely matching experimental stoichiometries (see Table \ref{tab:all-comp}). The obtained SWy-1 model recapitulates the experimentally determined Wyoming montmorillonite (SWy-1/SWy-2/SWy-3) composition substantially better than previously used simplified models (SWy-simp.).\cite{Underwood2016, Underwood2015}
Both the KGa-1 and IMt-1 models feature a slightly higher negative charge than their reference structures. We note that these reference structures feature small inclusion of \ce{Ti^{4+}} and trace amounts of \ce{Mn^{2+}}. Since no ClayFF force field parameters are currently available for these elements, these are not included in the \textit{ClayCode}-produced structures.
For \ce{Ti^{4+}}, we note that the experimental assignment in the reference structures may not always correspond to an isomorphic substitution. Both \citeauthor{Dolcater1970} and \citeauthor{Shoval2008StudySpectroscopy} have observed strong bands of accessory anatase, \ce{TiO2}, in Raman spectra of kaolinites.\cite{Dolcater1970, Shoval2008StudySpectroscopy} 
Specifically, \citeauthor{Dolcater1970} found that \SI{86}{\percent} of \ce{Ti^{4+}} in kaolinites is incorporated in the form of anatase, with \citeauthor{Shoval2008StudySpectroscopy} also noting minor incorporation of \ce{Ti^{4+}} in the form of octahedral substitutions which are not expected to exceed \SI{0.02}{atoms\per\uc}.\cite{Shoval2008StudySpectroscopy}
\textit{ClayCode} handling of these minor substitutions resulted in slightly elevated charges for IMt-1 and KGa-1 models with respect to their experimental counterparts.

\begin{figure}[H]
    \includegraphics[width=\textwidth]{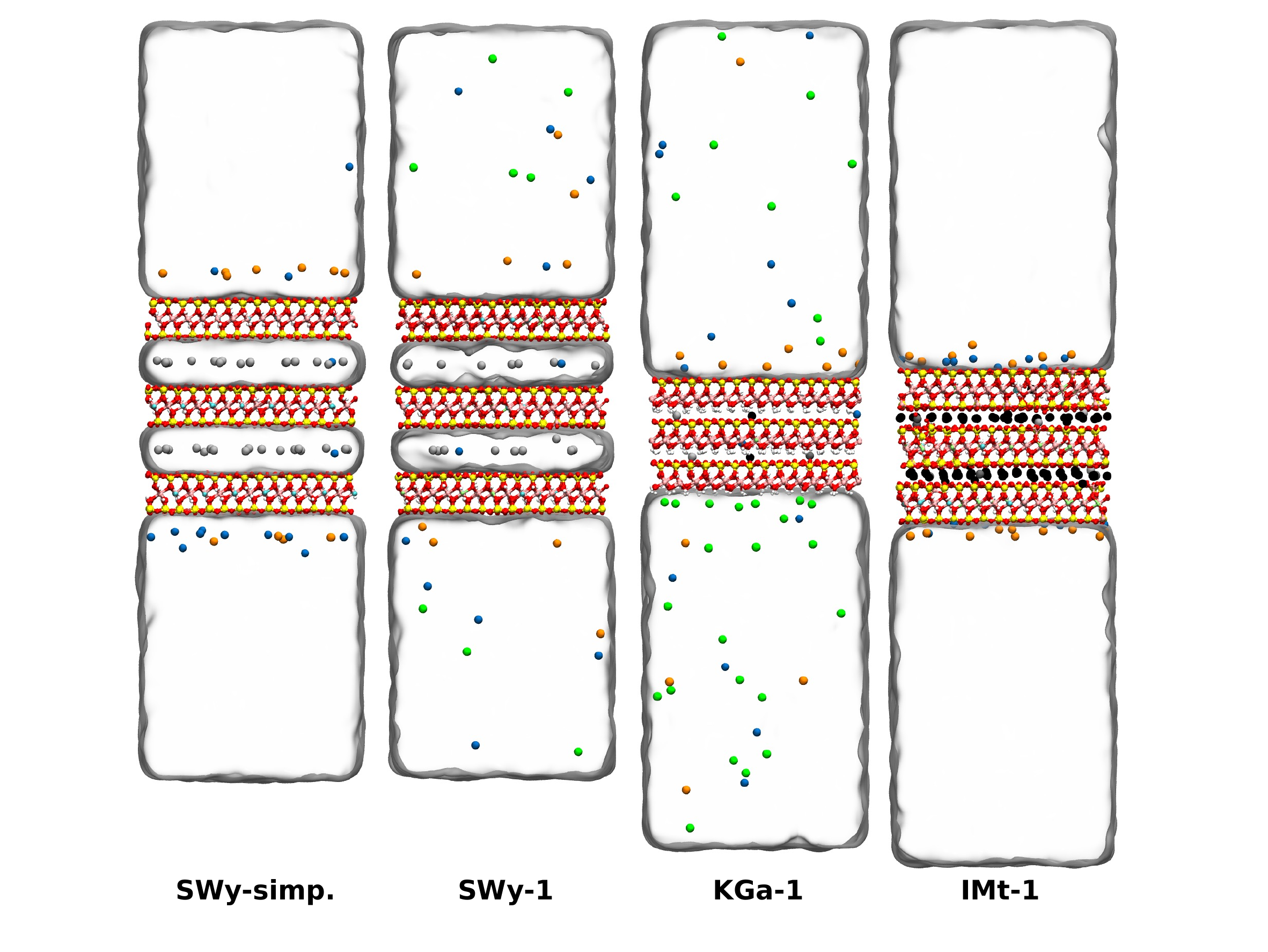}
    \caption{Side view of the four systems produced with \textit{ClayCode}. Clay layers in the middle, water is represented as transparent surfaces and ionic species are shown as spheres of different colour.}
    \label{fig:clay_structures}
\end{figure}


\subsection{Cation adsorption on two Wyoming montmorillonite models}

To investigate the importance of using accurate clay models in simulations, we compared \ce{Na+} and \ce{Ba^{2+}} adsorption onto the simplified Wyoming montmorillonite model (SWy-simp.), against the more realistic model (SWy-1) set up with \textit{ClayCode} (see Table \ref{tab:all-comp}). 

\begin{figure}[H]
    \includegraphics[width=\textwidth]{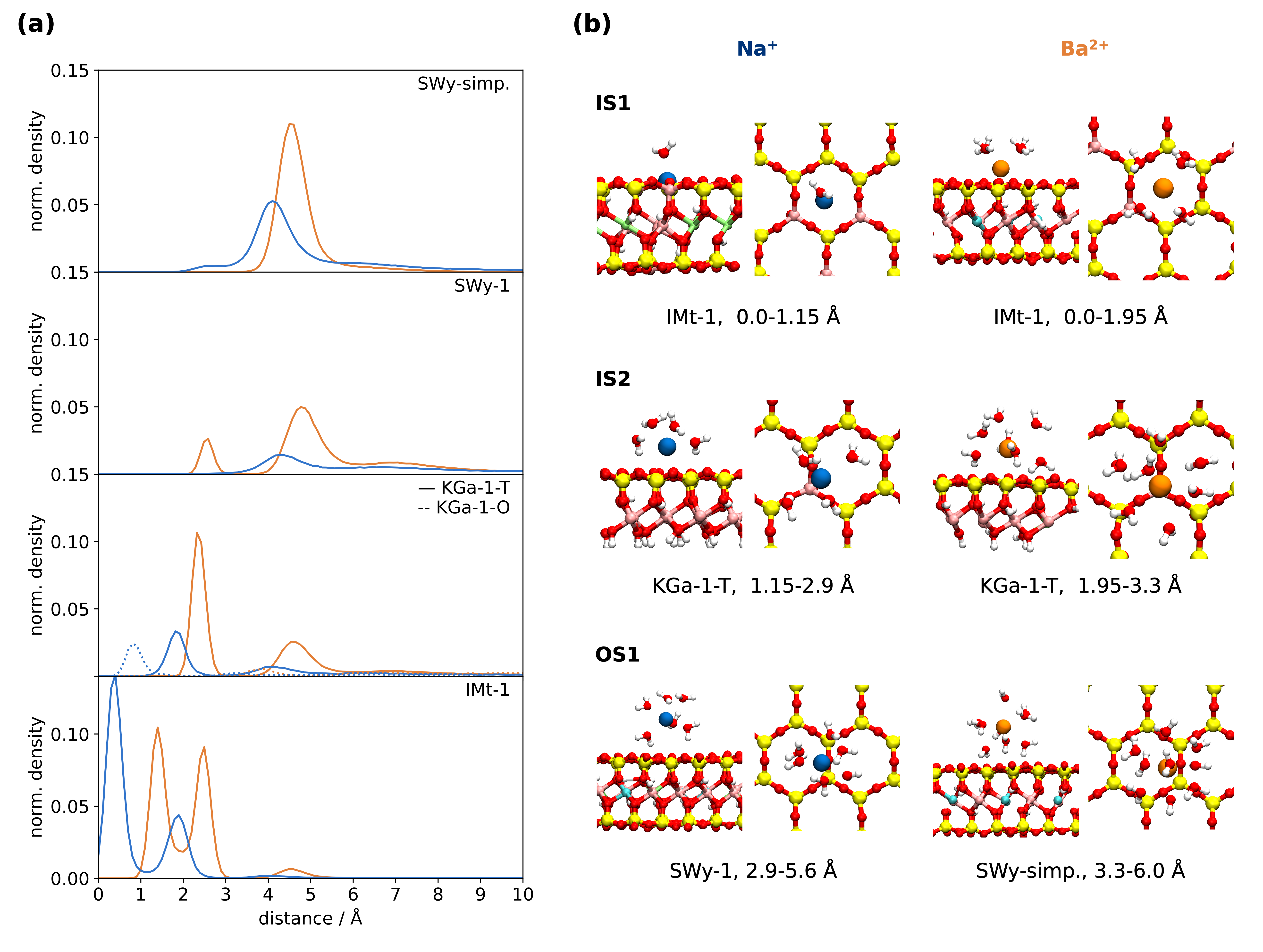}
    \caption{(a) Normalized linear densities of \ce{Na+} (blue) and \ce{Ba^{2+}} (orange) ions for all simulated systems. The $x$-axis represents the distance of ions from the closest clay distribution peak (see Figure \ref{fig:S1}). (b) For both \ce{Na+} and \ce{Ba^{2+}} ions, representative snapshots of inner-shell (IS1, IS2) and outer-shell (OS1) adsorption modes. Water molecules in the first hydration shell of each ion are shown. Snapshots representing each peak in every simulation are presented in Figure \ref{fig:S3}.}
    \label{fig:ads_render}
\end{figure}

Comparing the two models, we find that the SWy-simp. features substantially higher adsorption for both \ce{Na+} and \ce{Ba^{2+}} (see Figure \ref{fig:ads_render} a and Table \ref{tab:ads_cap}). Specifically, SWy-simp. adsorbs \SI{80}{\percent} of total \ce{Na+}, while SWy-1 adsorbs significantly less -- only \SI{36}{\percent}. Similarly, for \ce{Ba^{2+}}, SWy-simp. adsorbs nearly all cations present (\SI{99.5}{\percent}), while SWy-1 adsorbs a slightly smaller amount (\SI{82}{\percent}). This difference is consistent with the difference in total charge between these two models (-1.0 \si{\elementarycharge\per\uc} for SWy-simp. and = -0.54 \si{\elementarycharge\per\uc} for SWy-1).

These differences also mean that the selectivity for \ce{Ba^{2+}} over \ce{Na+} is very different for these two clay models. SWy-simp. shows only a slight preference for \ce{Ba^{2+}} over \ce{Na^{+}} (1.25:1), while the realistic SWy-1 model is much more selective (2.2:1). The interactions of these cations with SWy-simp. were previously studied with metadynamics by \citeauthor{Underwood2016}. The study highlighted that unbiased MD was able to reproduce well the adsorption profiles for these two cations, and calculated their exchange equilibrium constant of 1.33, in agreement with the selectivity ratio predicted in this work for the same simplified montmorillonite model.\cite{Underwood2016}

There exist many studies determining experimental exchange equilibrium constants for montmorillonites/bentonites, owing to their applications as barriers for various heavy metal wastes. Older works report exchange constants between 1.1 and 2.3 for montmorillonites, though notably for slightly different varieties to the one in this work.\cite{Benson1982, Lewis1963} 
A more recent study by \citeauthor{Klinkenberg} has found that the exchange coefficients of \ce{Ba^{2+}} for \ce{Na+} on Wyoming montmorillonite (SWy-1) to be concentration-dependent, increasing with increased ionic strength (2.46 for 0.3 M solution, 2.0 for under 0.02 M).\cite{Klinkenberg} This, again, is in perfect agreement with the findings from our simulation of the realistic SWy-1 clay.

The higher preference for divalent cation by the realistic SWy-1 can be attributed to the formation of the \ac{IS} complex in addition to the \ac{OS} complex, which is also observed for in the SWy-simp. model (peaks at \SI{2.5}{\angstrom} and \SI{4.5}{\angstrom} distance away from the surface, respectively, see Figure \ref{fig:ads_render}). The IS complex of \ce{Ba^{2+}} is facilitated by the small number of Al substitutions in the \ac{T} sheet of SWy-1.
Interestingly, \citeauthor{Zhang2001adsorption} observed via EXAFS that at high pH (i.e., not on the edge-site) and high concentrations, \ce{Ba^{2+}} forms both \ac{IS}and \ac{OS} complexes on SWy-1 clay.\cite{Zhang2001adsorption}

Our comparison of cation adsorption on two models of Wyoming montmorillonite -- one simplified used in previous works and one realistic assembled with \textit{ClayCode} -- highlights the importance of accounting for the finest structure features in clay minerals to truly gain atomic-level insight into experimental observables.

\begin{table}[]
\centering
\caption{Ionic adsorption capacities on siloxane surface only, split into the adsorption type determined by the distance away from the surface, and selectivity for \ce{Ba^{2+}} over \ce{Na+} for each clay system modelled. For representative structures see Figure \ref{fig:S3}.}
\label{tab:ads_cap}
\resizebox{\columnwidth}{!}{%
\begin{tabular}{c|cccccccccc|c}
 &
  \multicolumn{10}{c|}{Ion adsorption per adsorption type (\%)} &
  \multirow{3}{*}{\begin{tabular}[c]{@{}c@{}}Selectivity\\ \ce{Ba^{2+}}:\ce{Na+}\end{tabular}} \\ \cline{2-11}
 &
  \multicolumn{5}{c|}{\ce{Na^+}} &
  \multicolumn{5}{c|}{\ce{Ba^{2+}}} &
   \\ \cline{2-11}
\begin{tabular}[c]{@{}c@{}}Adsorption\\ (distance, Å)\end{tabular} &
  \begin{tabular}[c]{@{}c@{}}IS1\\ (0.0-1.15)\end{tabular} &
  \begin{tabular}[c]{@{}c@{}}IS2\\ (1.15-2.9)\end{tabular} &
  \begin{tabular}[c]{@{}c@{}}OS1\\ (2.9-5.6)\end{tabular} &
  \begin{tabular}[c]{@{}c@{}}OS2\\ (5.6-10)\end{tabular} &
  \multicolumn{1}{c|}{\textit{not ads.}} &
  \begin{tabular}[c]{@{}c@{}}IS1\\ (0.0-1.95)\end{tabular} &
  \begin{tabular}[c]{@{}c@{}}IS2\\ (1.95-3.3)\end{tabular} &
  \begin{tabular}[c]{@{}c@{}}OS1\\ (3.3-6.0)\end{tabular} &
  \begin{tabular}[c]{@{}c@{}}OS2\\ (6.0-10)\end{tabular} &
  \textit{not ads.} &
   \\ \hline
SWy-simp. &
  - &
  3.5 &
  59.8 &
  16.4 &
  \multicolumn{1}{c|}{\textit{20.3}} &
  - &
  0.1 &
  94.0 &
  5.4 &
  \textit{0.5} &
  1.25 \\
SWy-1 &
  - &
  0.3 &
  18.9 &
  16.7 &
  \multicolumn{1}{c|}{\textit{64.1}} &
  - &
  10.1 &
  50.1 &
  21.7 &
  \textit{18.1} &
  2.2 \\
KGa-1-T &
  0.1 &
  18.0 &
  9.2 &
  7.4 &
  \multicolumn{1}{c|}{\textit{65.2}} &
  0.2 &
  42.7 &
  25.2 &
  10.0 &
  \textit{21.9} &
  2.2 \\
IMt-1 &
  65.2 &
  26.7 &
  2.2 &
  1.1 &
  \multicolumn{1}{c|}{\textit{4.8}} &
  49.2 &
  43.3 &
  5.3 &
  0.7 &
  \textit{1.5} &
  1.1
\end{tabular}%
}
\end{table}


\subsection{Interaction of cations with realistic montmorillonite, illite, and kaolinite models}

In addition to Wyoming montmorillonite, we also investigated the effect of clay composition and structure for \ce{Na+} and \ce{Ba^2+} adsorption on kaolinite KGa-1 and illite IMt-1 clays. Alike montmorilonite, illite is a 2:1 clay exposing two tetrahedral surfaces to the solvent. On the other hand, kaolinite is a 1:1 clay exposing one tetrahedral and one octahedral surface to the solvent. 
We first examine ion adsorption on the two KGa-1 surfaces.
We note that experimentally it is near-impossible to distinguish what kaolinite surface participates in the adsorption, or to quantify the ratio of the exposed surfaces in the solution. 
However, with simulations we can easily separate these two surfaces. Figure \ref{fig:ads_render} compares the adsorption of cations on the octahedral (KGa-1-O, dashed line) and on the tetrahedral (KGa-1-T, solid line) surfaces, showing that only minimal \ce{Na+} and no \ce{Ba^2+} are adsorbed onto the KGa-1-O. For this reason, we omit the contribution of the octahedral surface from the adsorption capacity calculations presented in Table \ref{tab:ads_cap}.

The behaviour of the KGa-1 is similar to the realistic SWy-1 model, with the same adsorption preference for \ce{Ba^{2+}} over \ce{Na+} of 2.2. Similarly, the adsorption of both cations is found above the tetrahedral Al substitution (see IS2 renderings on Figure \ref{fig:ads_render} b), which is slightly higher in KGa-1 than in SWy-1 (KGa-1 features 0.17 Al per 4 tetrahedral positions, SWy-1 only 0.03 Al per 8, see Table \ref{tab:all-comp}). 

For IMt-1, nearly all of the \ce{Na+} and \ce{Ba^{2+}} ions were adsorbed, always forming \ac{IS} complexes. In fact, we identified two IS complexes: IS2 at approximately \SI{2.0}{\angstrom} for \ce{Na+} and \SI{2.5}{\angstrom}  for \ce{Ba^{2+}} distance from the surface and IS1 much closer to the surface at just \SI{0.5}{\angstrom} for \ce{Na+} and \SI{1.5}{\angstrom} for \ce{Ba^{2+}} (Figure \ref{fig:ads_render}). While IS2 is identical to the IS adsorption mode found on KGa-1 and SWy-1, where the cation sits directly above the tetrahedral Al substitution, IS1 is facilitated by the frequent proximate Al substitutions that act like pliers, holding the cation tightly between them (see IS1 on Figure \ref{fig:ads_render}). In the case of the smaller \ce{Na+} (ionic radius of \SI{1.16}{\angstrom} compared to \SI{1.49}{\angstrom} for \ce{Ba^{2+}}), ions nearly fully submerge into the silicate crown of the tetrahedral surface, with only one water coordinated above them. \ce{Ba^{2+}} protrudes further, allowing for four waters to sit above it.
While the near-complete adsorption of cations on IMt-1 does not enable an accurate estimation of selectivity, in line with expectations, we observe a slightly higher preference for \ce{Ba^{2+}}.

Overall, our simulations, in agreement with experimental works, have shown that all modelled clays have a slight preference for \ce{Ba^{2+}}, resulting in over 70\% its removal from solution.\cite{Atun2003adsorption, Klinkenberg} Furthermore, leveraging accurate molecular models allowed us to gain atomistic insights on the specific adsorption mechanisms for each clay, and highlight how isomorphic substitutions determine their specific chemophysical properties.


\section {Conclusions}

In this work we introduce \emph{ClayCode} -- a comprehensive and user-friendly software designed to facilitate and advance molecular modelling of clay materials. 
By enabling the construction of clay models that closely resemble their experimentally determined structures, \emph{ClayCode} addresses the critical need for realistic and representative simulations that are vital for interpreting and predicting the behaviour of clay minerals in varied scientific and engineering contexts.

Through the simulation of \ce{Na+} and \ce{Ba^{2+}} ions adsorption on common clay minerals -- Wyoming montmorillonite (both simplified and realistic models), Georgia kaolinite and Montana illite -- we demonstrated that the structural accuracy of the models has a significant effect on simulation outcomes. 
For instance, our findings illustrate that a realistic representation of montmorillonite predicts cation adsorption with greater fidelity to experimental results, thereby revealing essential details about ion exchange dynamics that were not captured by simpler, idealised models.
Another outcome enabled by setting up clay models with \emph{ClayCode}, was the variability in ion adsorption efficiency and selectivity among different types of clays, all driven by subtle differences in their compositional and structural properties.
This further reinforces the importance of employing accurate clay models for studies where molecular-level interactions govern the macroscopic properties.

Looking forward, while \emph{ClayCode} has already proven its utility in modelling planar, hydrated clay systems, its flexible and modular design lays a robust foundation for future enhancements.
The goals of forthcoming developments are to extend its capabilities to include pH-dependent edge sites and non-planar geometries.
Additionally, further development of an integrated analysis module within \emph{ClayCode} will streamline the workflow from model construction to analysis, facilitating rapid and in-depth interpretation of simulation results.

In conclusion, \emph{ClayCode} represents a significant step forward in computational clay mineral modelling, enabling researchers to build more accurate models that can better mimic real-world materials. This, in turn, enhances the reliability of simulations and the insights they can provide, supporting more informed decisions in the fields of environmental and material sciences.

\section {Data Availability}

\emph{ClayCode} is available open-source at \href{https://github.com/Erastova-group/ClayCode}{github.com/Erastova-group/ClayCode}, 
DOI: \href{http://doi.org/10.5281/zenodo.11219451}{10.5281/zenodo.11219451}.
A workshop material dedicated to ClayCode is also available: \href{https://github.com/Erastova-group/ClayCode-workshop}{github.com/Erastova-group/ClayCode-workshop}.
Manual and Tutorials are available at \href{https://claycode.readthedocs.io}{claycode.readthedocs.io}.

\begin{acknowledgement}
This work made use of the facilities of the N8 Centre of Excellence in Computationally Intensive Research (N8 CIR) provided and funded by the N8 research partnership and EPSRC (Grant No. EP/T022167/1). The Centre is coordinated by the Universities of Durham, Manchester, and York.
M.T.D. acknowledges the support of the Engineering and Physical Sciences Research Council (Grant EP/P016499/1).
V.E. acknowledges the support of a Chancellor's Fellowship by the University of Edinburgh for herself and H.P.
The authors thank Sarah Stewart for her help in setting up the online documentation for ClayCode. 
\end{acknowledgement}

\bibliography{references}

\newpage


\renewcommand{\theequation}{S\arabic{equation}}
\renewcommand{\thetable}{S\arabic{table}}
\renewcommand{\thefigure}{S\arabic{figure}}
\setcounter{equation}{0}
\setcounter{table}{0}
\setcounter{figure}{0}
\setcounter{page}{1}

\begin{suppinfo}


\textbf{Input files}

\noindent YAML files - user inputs build specification for each clay type:
\begin{itemize}
    \item \texttt{SWy-simplified.yaml}
    \item \texttt{SWy-1.yaml}
    \item \texttt{KGa-1.yaml}
    \item \texttt{IMt-1.yaml}
\end{itemize}

\noindent CSV file - Experimentally determined structure file: 
\texttt{exp\_clay.csv}

\noindent These files are also available for download from {}\texttt{/paper/} directory within \emph{ClayCode} at\\ \href{https://github.com/Erastova-group/ClayCode}{github.com/Erastova-group/ClayCode} 
\\

\begin{table}[H]
\caption{Simulation box dimensions after equilibration, number and type of inserted bulk ions, and \ac{IL} and bulk solvent molecules for the simplified (SWy-simp.) and realistic SWy-1, IMt-1, and KGa-1 models.}
\label{tab:ion_numbers}
\begin{tabular}{c|ccc|ccc|cc}
\multirow{2}{*}{Clay} & \multicolumn{3}{c|}{Sim. box size (nm)} & \multicolumn{3}{c|}{Bulk ions} & \multicolumn{2}{c}{Water molecules} \\ \cline{2-9} 
                   & \textit{x} & \textit{y} & \textit{z} & \ce{Na+} & \ce{Ba^{2+}} & \ce{Cl-} & IL & bulk \\ \hline
SWy-simp. & 3.625         & 4.50         & 12.91        & 13  & 11   & 0   & 690        & 3557 \\
SWy-1       & 3.63         & 4.51         & 12.80        & 9  & 9    & 8   & 590        & 3369 \\
IMt-1       & 3.63         & 4.51         & 14.37        & 20  & 19   & 0   & --         & 4270 \\
KGa-1       & 3.62         & 4.49         & 14.10        & 12  & 12   & 30  & --         & 4228 \\
\end{tabular}
\end{table}


\begin{figure}
\centering
\includegraphics[width=0.75\linewidth]{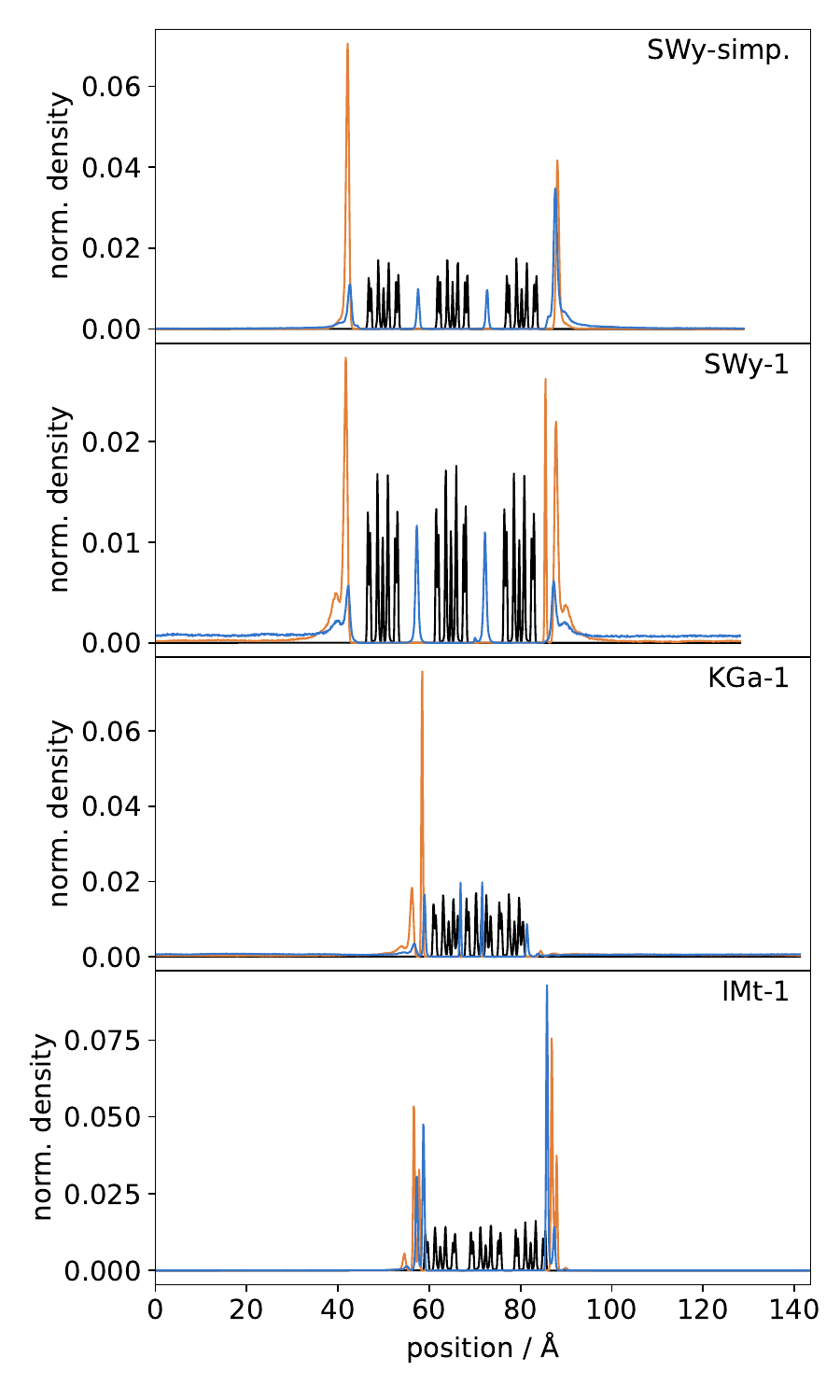}
\caption{Raw normalized densities for $Na^{+}$ (blue) and $Ba^{2+}$ (orange) and clay (black) atoms. The $x$-axis represents absolute position in the $z$-axis of each simulation box.}
\label{fig:S1}
\end{figure}


\begin{figure}
\centering
\includegraphics[width=0.75\linewidth]{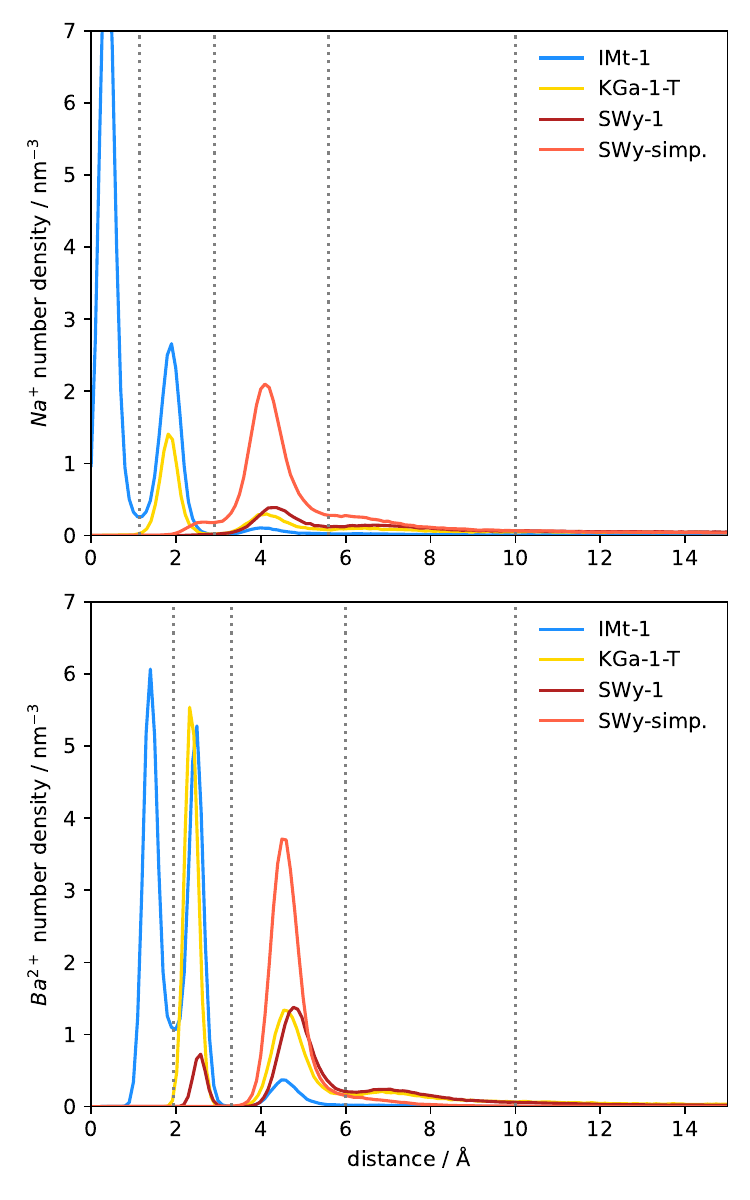}
\caption{Number density of $Na^{+}$ (above) and $Ba^{2+}$ (below) ions exposed to different surfaces. Gray dashed lines represent the cutoff distances adopted to separate the partial densities into regions associated with different adsorption modes (as listed in Table \ref{tab:ads_cap}). For \ce{Na+}: IS1 is at 0.00 - 1.15 \AA 
~away from the surface, IS2 at 1.15 - 2.9 \AA, OS1 2.9 - 5.6 \AA, OS2 5.6 - 10.0 \AA. For \ce{Ba^{2+}}: IS1 is at 0.0 - 1.95 \AA, IS2 at 1.95 - 3.3 \AA, OS1 at 3.3 - 6.0 \AA, OS2 at 6.0 - 10.0 \AA. Beyond 10 \AA 
~distance away from the surface cations are in the bulk, i.e., not adsorbed.}
\label{fig:S2}
\end{figure}


\begin{figure}
\centering
\includegraphics[width=1\linewidth]{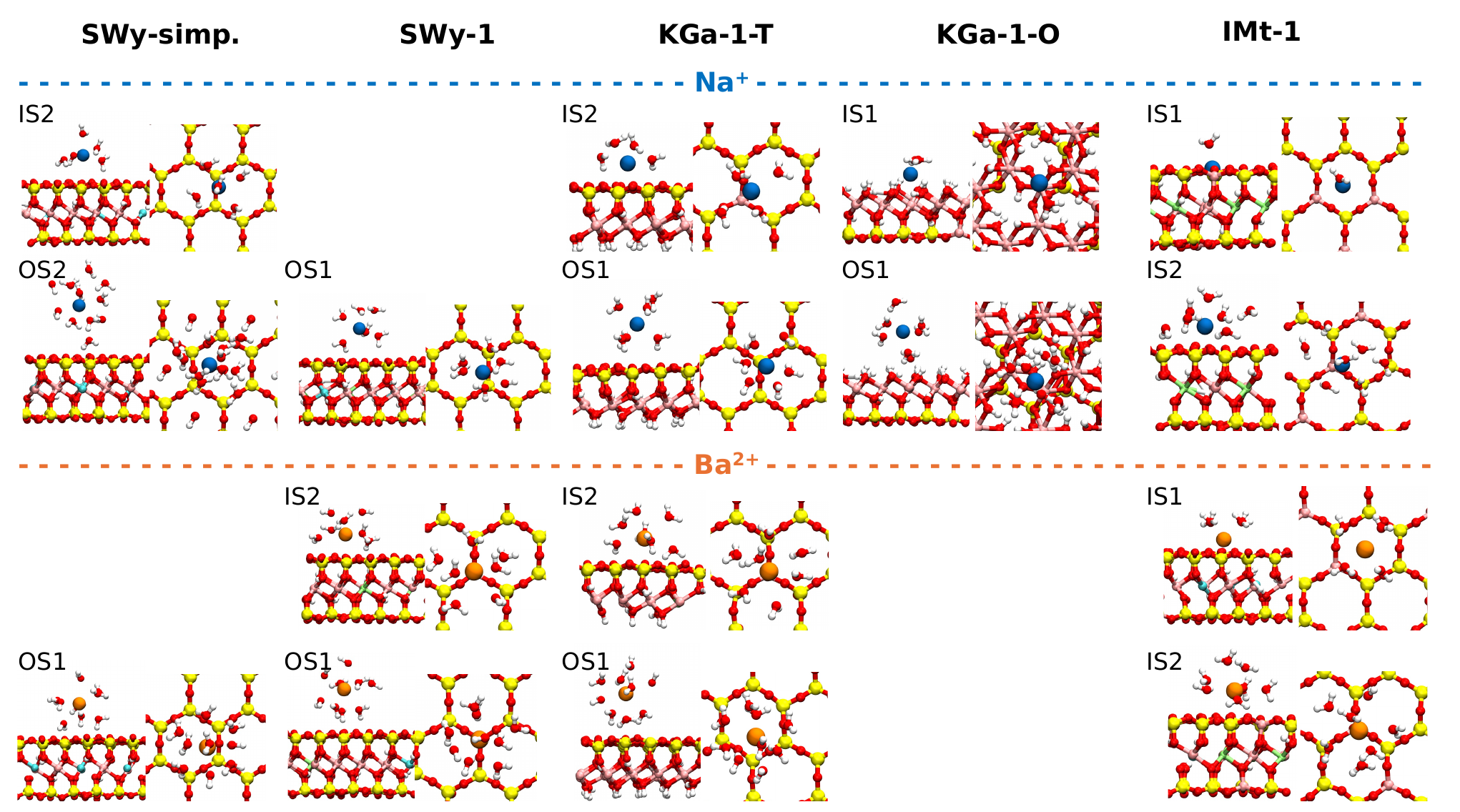}
\caption{Renderings of representative mechanisms of adsorption for \ce{Na+} and \ce{Ba^{2+}} cations on the clays. Only the water molecules forming hydration shell around adsorbed ions are shown. The adsorption mechanisms are defined by the linear density profiles (Fig. \ref{fig:S2}) and are as follows, for \ce{Na+}: IS1 is at 0.00 - 1.15 \AA 
~away from the surface, IS2 at 1.15 - 2.9 \AA, OS1 2.9 - 5.6 \AA, OS2 5.6 - 10.0 \AA;  for \ce{Ba^{2+}}: IS1 is at 0.0 - 1.95 \AA, IS2 at 1.95 - 3.3 \AA, OS1 at 3.3 - 6.0 \AA, OS2 at 6.0 - 10.0 \AA.}
\label{fig:S3}
\end{figure}

\end{suppinfo}

\end{document}